\documentclass[
reprint,
superscriptaddress,
showpacs,
preprintnumbers,
nofootinbib,
amsmath,
amssymb, 
aps,
prd,
dvipsnames,
floatfix
]{revtex4-2}
\pdfoutput=1

\usepackage{amsmath}
\usepackage{amsfonts}
\usepackage{amssymb}
\usepackage{mathrsfs}
\usepackage{slashed}
\usepackage{graphicx}
\usepackage[dvipsnames]{xcolor}
\usepackage{multirow,makecell}
\usepackage{listings}
\usepackage{upgreek}
\usepackage{bm}
\usepackage[normalem]{ulem}
\usepackage{booktabs}
\usepackage{xspace}
\usepackage{physics}
\usepackage{adjustbox}
\usepackage{chemformula}
\usepackage[inline]{enumitem}
\usepackage{fontawesome5} % for code link icons
\usepackage[colorlinks=tr   ue,linkcolor=purple,citecolor=purple,allcolors=purple]{hyperref}
\usepackage[capitalise]{cleveref}

\makeatletter
\AddToHook{cmd/appendix/before}{\def\cref@section@alias{appendix}}
\makeatother
\definecolor{orcidlogocol}{HTML}{A6CE39}
\usepackage{isotope}

\definecolor{brightpink}{rgb}{1.0, 0.0, 0.5}
\definecolor{emerald}{rgb}{0.31, 0.78, 0.47}
\definecolor{limegreen}{rgb}{0.2, 0.8, 0.2}
\definecolor{blue-violet}{rgb}{0.33, 0.17, 0.89}

\usepackage{array}   % put this early in the preamble
\usepackage{makecell}
\usepackage{tabularx}
\newcolumntype{C}[1]{>{\centering\arraybackslash}m{#1}} % vertically centered, fixed width
\newcolumntype{P}[1]{>{\centering\arraybackslash}p{#1}} % top aligned, fixed width

\definecolor{MH}{rgb}{0.0,0.6,0.9}

\newcommand{\myorcid}[1]{\href{https://orcid.org/#1}{\textcolor{orcidlogocol}{\faOrcid} #1}}
\newcommand{\JSNSsquared}{JSNS$^2$\xspace}

\begin{document}

\title{Long-Lived Particles from Meson and Muon Decays at Rest at Spallation Sources}

\newcommand{\IFT}
{Instituto de Física Teórica UAM-CSIC, Calle Nicolás Cabrera 13-15, Cantoblanco E-28049 Madrid, Spain}

\newcommand{\IJCLAB}
{IJCLab, Pôle Théorie (Bat. 210), CNRS/IN2P3, 91405 Orsay, France
}

\author{Matheus Hostert}
\email[]{matheus-hostert@uiowa.edu}
\thanks{\myorcid{0000-0002-9584-8877}}
\affiliation{Department of Physics and Astronomy, University of Iowa, Iowa City, IA 52242, USA}

\author{Salvador Urrea}
\email[]{salvador.urrea@ijclab.in2p3.fr}
\thanks{\myorcid{0000-0002-7670-232X}}
\affiliation{\IJCLAB}

\date{\today}

\begin{abstract}
We provide a systematic survey of spallation sources around the world and their potential to search for new light particles.
We study the sensitivity of existing neutrino detectors to the decay in flight of light particles produced in the decay at rest of pions, muons, and kaons.
At J-PARC, we show that the magnetized gaseous argon detectors of ND280 could place leading limits on light particles decaying to $e^+e^-$ and that the liquid-scintillator detectors of the J-PARC Sterile Neutrino Search at the JSNS (JSNS$^2$) experiment can exploit double- and triple-coincidence signals from $\mu^+\mu^-$ and $\pi^+\pi^-$ to place new limits on particles produced in kaon decay.
We find that the Coherent Captain Mills detector at Los Alamos and the suite of COHERENT detectors at Oak Ridge, despite their smaller size, would also have a promising sensitivity to particles produced in pion and muon decays, depending on background levels.
Spallation sources have the potential to explore more than an order of magnitude beyond current constraints in some new physics models, encouraging further study on data acquisition and background rejection by experimental collaborations.
\end{abstract}

\maketitle

% TABLE OF CONTENTS
{
% disable subsections and subsubsections in the TOC
\makeatletter
\def\l@subsubsection#1#2{}
\makeatother

\hypersetup{linkcolor=black}
\tableofcontents
} % End of TOC block

%%%%%%%%%%%%%%%%%%%%%%%%%%%%%
\section{Introduction}
\label{sec:intro}

Searches for light and feebly interacting particles provide a compelling direction in the pursuit of physics beyond the Standard Model (SM)~\cite{Antel:2023hkf}.  
These new particles can have important implications for open problems in particle physics, like the origin of neutrino masses, the Higgs hierarchy problem, and dark matter.  
As such, they provide an exciting opportunity for discovery.  
Their couplings to the SM must be small to have evaded detection thus far, inviting the idea that they could be part of a ``dark'' or ``hidden'' sector of particles.  
When searching for dark sectors in the laboratory, neutrino experiments play a crucial role~\cite{Batell:2022xau}.  
Neutrino detectors are designed to have large volumes and low background rates, and are usually exposed to neutrino sources from high-intensity proton beams.  
A detailed reconstruction of neutrino interaction products is crucial for background rejection and is achieved through a variety of techniques, including the collection of Cherenkov or scintillation light as well as charge collection in high- or low-density active materials.  
As a result, neutrino facilities are well suited to look for rare processes associated with the decay or scattering of new light particles.  
In many cases, they offer the only avenue to search for particles that interact with the SM through interactions weaker than the SM Weak force.

In well-motivated dark sector models, it is common to find new states that decay to SM particles with lifetimes on macroscopic scales.  
These long-lived particles (LLPs) can be produced in rare decays of pions, kaons, and muons, and can subsequently propagate to large-volume detectors located tens to hundreds of meters away from their origin.  
Their experimental signature is then characterized by an energy deposition in the detector in the form of pairs of charged particles or photons, like $e^+e^-$ or $\gamma\gamma$.  
We refer to these as long-lived particle decay-in-flight (LLP-DIF) signatures.  
In many models, the LLP's energy is fully converted into visible final states, resulting in a signal that may not only be a bump in the energy spectra, but also typically more energetic than potential neutrino- or neutron-induced backgrounds.

In this article, we explore the sensitivity of neutrino detectors at spallation neutron sources to LLPs.  
As we will see, the combination of the large number of protons on target (POT) and large-volume neutrino detectors makes these facilities particularly well suited for LLP searches.  
Theoretical targets of LLPs include heavy neutral leptons (HNLs), axion-like-particles (ALPs), and other mediator particles like long-lived scalar bosons.
While some experiments like the Coherent Captain-Mills experiment~\cite{CCM:2021yzc,CCM:2021leg} and the COHERENT experiment~\cite{Barbeau:2021exu,Akimov:2022oyb} have searched for light dark matter scattering signatures, none so far has conducted a dedicated search for LLP-DIF signatures.
This is despite the fact that DIF signals can be, in some cases, far easier to pick out than neutrino or dark matter scattering events due to their higher energy and monochromatic nature.

The potential of meson and muon decays as probes of physics beyond the SM at spallation facilities has long been recognized.  
Data from the Liquid Scintillator Neutrino Detector (LSND) at the Los Alamos Meson Physics Facility (LAMPF)~\cite{LSND:1996ubh,LSND:2001akn} has been widely used to constrain some models of light dark matter and LLPs~\cite{Batell:2009di,Essig:2010gu,deNiverville:2011it,deNiverville:2015mwa,Foroughi-Abari:2020gju}.\footnote{Most limits are derived using the measurement of elastic neutrino-electron scattering ($\nu \,e^-\to \nu \, e^-$) which is a different sample from the anomalous inverse beta decay (IBD) events ($\overline\nu_e p^+ \to e^+ n$) observed by the experiment~\cite{LSND:1996ubh}.}  
However, many dark sector models remain untested by LSND due to limitations on either the detector, exposure, or on the analysis technique.
Here, we address some of these shortcomings by considering data from the KArlsruhe Rutherford Medium Energy Neutrino (KARMEN) experiment~\cite{KARMEN:1998xmo}.
We note that KARMEN has been largely neglected in past literature due to its slightly smaller exposure and detector, however, in some models it can still place competitive or leading limits thanks to the more inclusive nature of the measurements.
Understanding the current exclusions by LSND and KARMEN data is crucial to delineate the regions of parameter space where the greatest potential for future discoveries lies.

Previous theoretical work has also pointed out several new possible experimental signatures, including HNLs and ALPs at the JSNS$^2$ experiment at J-PARC~\cite{Ema:2023buz,Ema:2023tjg}, dark matter scattering from neutral meson decays in flight~\cite{deNiverville:2015mwa,COHERENT:2021pvd} or charged meson decay at rest~\cite{Dutta:2020vop,Dutta:2023fnl,CCM:2023itc}, neutron-induced production of exotic mediators~\cite{Dutta:2024yjp}, and production of new particles in neutrino–nucleus interactions~\cite{Bolton:2021pey,Alonso-Gonzalez:2023tgm,Candela:2023rvt,Schneider:2024eej,Candela:2024ljb}.
Similar strategies are under consideration for future facilities, for example the SHiNESS detector~\cite{Abele:2022iml,Soleti:2023hlr} at the planned European Spallation Source (ESS)~\cite{Garoby:2017vew}.
However, despite this growing number of applications, there still remain several low-hanging fruits, i.e., minimal and testable theoretical models, that modern detectors could target.  
Here, we provide a systematic approach to finding those, focusing primarily on LLPs produced in the decay at rest of pions, muons, and kaons.
We identify the most minimal scenarios that cover a variety of signatures, paving the way for future experimental analyses at modern facilities.

This article is divided as follows.
We begin with a review of the neutrino and LLP production at spallation sources in \cref{sec:spallation}.
We then present an overview of the sources  around the world and the neutrino detectors currently in operation in \cref{sec:sources}.
We focus on the most relevant sources for LLP searches: the Spallation Neutron Source (SNS) at Oak Ridge National Laboratory, the Japanese Spallation Neutron Source (JSNS) at J-PARC, and the Los Alamos Neutron Science Center (LANSCE).
We also introduce the neutrino detectors discussing background mitigation strategies and the advantages and disadvantages of each detector technology in the context of LLP signatures.
In Sections~\ref{sec:hnls}, \ref{sec:scalars}, and \ref{sec:ALPs}, we introduce specific dark sector models with LLPs and discuss their predictions for DIF signatures, presenting projected experimental sensitivities in comparison with current bounds.  
Finally, we conclude in \cref{sec:conclusions}.

%%%%%%%%%%%%%%%%%%%%%%%%%%%%%%%%%%%%%%%%%%
\section{Spallation Sources}
\label{sec:spallation}

Spallation sources are characterized by their high-intensity proton beams of $\mathcal{O}(1)$~GeV energies impinging on high-$Z$ targets.
They provide an intense and controlled source of neutrons for a variety of applications in materials science or fundamental physics.
For reviews on the physics of spallation sources, see, e.g., Refs.~\cite{Filges2009,BAUER2001505,David:2015ura}.
These sources are typically located at large facilities, such as the SNS at Oak Ridge National Laboratory, the JSNS at J-PARC, and the LANSCE at the Los Alamos National Laboratory.
In addition to neutrons, they also produce a large number of light mesons, such as pions and kaons, which quickly lose energy within the target and decay at rest (DAR) to neutrinos and muons.

\subsection{Spallation} 

In the context of spallation neutron sources, a spallation reaction is the process by which a particle, usually a proton in the beam, strikes a nucleus with enough energy to excite it and knock out some of its constituents, creating a variety of lighter isotopes, mesons, and radiation in the process. 
This reaction typically develops in three main steps.
The first is the intra-nuclear cascade, where the incident proton collides with individual nucleons inside the nucleus, knocking some of them out and, in a fraction of these collisions, creating pions or kaons.
At this stage, the de Broglie wavelength of the incoming GeV protons can be much smaller than the nuclear radius, $\lambda_{p} \sim 0.1$~fm $<$ $ R_A \sim 1.2 A^{1/3}$~fm, indicating that the reaction can be accurately described by individual nucleon-nucleon collisions.
As the projectile proton exits the nucleus, the second step begins, where the energy injected is redistributed through secondary collisions within the nuclear volume and the nuclear target is left in a highly excited state.
In the third and final step called evaporation, the nucleus de-excites by emitting several lower-energy neutrons and other lighter fragments.
This sequence of events results in a large number of neutrons emitted per spalled nucleus.
For instance, a $0.8$~GeV proton collision on a Hg target can produce of the order of $10 - 20$ neutrons per incident proton depending on the thickness of the target.

Light meson production in the intra-nuclear cascade proceeds via resonant and non-resonant inelastic nucleon-nucleon scattering.
The most common mesons produced in spallation sources are pions, etas, and kaons, which can be produced, for instance, in the following reactions:
\begin{subequations}
\begin{align}
    pp  &\to p n \pi^+ (pp\pi^0)\,\, & (T_p \gtrsim 300~\text{MeV}) & \\
    pp  &\to p p \pi^+ \pi^- \,\, & (T_p \gtrsim 600~\text{MeV}) & \\
    pp  &\to p p \eta \,\, & (T_p \gtrsim 1.2~\text{GeV}) & \\
    pp  &\to p \Lambda^0 K^+ \,\, & (T_p \gtrsim 1.6~\text{GeV}) & \\
    pp  &\to p \Sigma^+ K^0 \,\, & (T_p > 1.8~\text{GeV}) & \\
    pp  &\to p p K^+ K^- \,\, & (T_p > 2.5~\text{GeV}) & 
\end{align}
\end{subequations}
where $T_p$ is the kinetic energy of the incident proton.
Once produced, $\pi^+$ and $K^+$ mesons quickly lose energy in the target and decay at rest.

\subsection{Neutrino production} 

Charged mesons initiate the main neutrino production chain via
\begin{align}
 &\pi^+/K^+  \,\to\, \mu^+ \, \nu_\mu, 
\end{align}
which produce monoenergetic neutrinos of $E_{\nu_{\mu}} \simeq 30\text{ MeV}$ and $236\text{ MeV}$, respectively.
The positively-charged muons also decay at rest,
\begin{align}
&\mu^+ \,\to\,  e^+ \, \nu_e  \, \overline\nu_\mu,
\end{align}
in this case to neutrinos with a continuum spectrum with endpoint of about $E_{\nu_e/\overline\nu_\mu} \lesssim 52\text{ MeV}$.
The vast majority of negatively-charged pions and kaons are quickly captured by the positively-charged nuclei in the target and constitute a much smaller fraction of the decays in the target.
For $\mu^-$, the nuclear capture also produces neutrinos that are approximately mono-energetic,
\begin{align}
\mu^- + {}_{Z}^{A}X \,\to\, \isotope[A][Z-1]{X} + \nu_\mu,
\end{align}
with $E_\nu \simeq 90 - 100\text{ MeV}$, depending on the nuclear target.
The particles that decay before stopping create a flux of higher-energy neutrinos focused in the same direction as the proton beam.
Neutrinos emitted in the $\beta$-decay of nuclei (or the few neutrons that decay before they are absorbed) are typically too low energy to be detected.
Neutrals such as $\pi^0$ and $\eta$ mesons quickly decay into photons before losing energy inside the target.

The neutrino flux can be used for a variety of applications, including the study of coherent neutrino-nucleus elastic scattering (CEvNS), namely $\nu + N_{\rm g.s.} \to \nu + N_{\rm g.s.}$.
This process was first measured on CsI in 2017 by the COHERENT collaboration~\cite{COHERENT:2017ipa,COHERENT:2021xmm} and has since been observed in Ar~\cite{COHERENT:2020ybo} and Ge~\cite{COHERENT:2024axu} targets.
CEvNS provides a powerful tool to study neutrino properties and constrain physics beyond the SM~\cite{Cadeddu:2023tkp}.
Neutrinos from $\mu$DAR also enable a measurement of the CC reaction $\nu_e + {\rm Ar} \to e^- + K^*$, which will be the primary detection channel at the DUNE far detector to extract the $\nu_e$ flux from an eventual galactic supernova~\cite{DUNE:2023rtr}.

\subsection{New Particle Production} 

The meson and muon decay in the target could also produce LLPs.
Such a source of LLPs would be particularly interesting because of its intensity and well-defined time profile.
The pulsed proton beam and short meson lifetimes mean that the LLP-DIF signatures, for example, follow a similar structure to the beam, delayed only by their time of flight to the respective detector location.
There are two main components for the neutrino flux.
The first is the prompt one from meson decays, where $\tau_{\pi^+}\sim 26$~ns and $\tau_{K^+} \sim 12$~ns.
This, in turn, can be followed by the delayed signal from muon decay, $\tau_\mu \sim 2.2~\mu$s. 
Depending on whether the LLP is produced by light mesons or muons, it can be associated with one or both of these time windows.
Having a signal that either ``beats" or ``lags behind" the fast neutrons emitted by the source is crucial to suppress beam-induced backgrounds.

For a fixed distance $L$ between the detector and the source, the delay in the time of flight between a new particle $X_{\rm LLP}$ and a neutrino emitted at the same time with the same momentum $p$, is given by
\begin{equation}\label{eq:time_delay}
\delta t \simeq L \frac{m_X^2}{2 p^2}.
\end{equation}
For instance, a particle of mass $m_X = 30$~MeV produced in $K \to \pi X_{\rm LLP}$ decays would arrive at a detector 48~m away after about $150$~ns with a relative time delay of approximately $\delta t \simeq 1$~ns.
For a heavier particle of $m_X = 300$~MeV, the time delay is much larger, $\delta t \simeq 300$~ns.

\begin{figure*}[t]
    \centering
    \includegraphics[width=\textwidth]{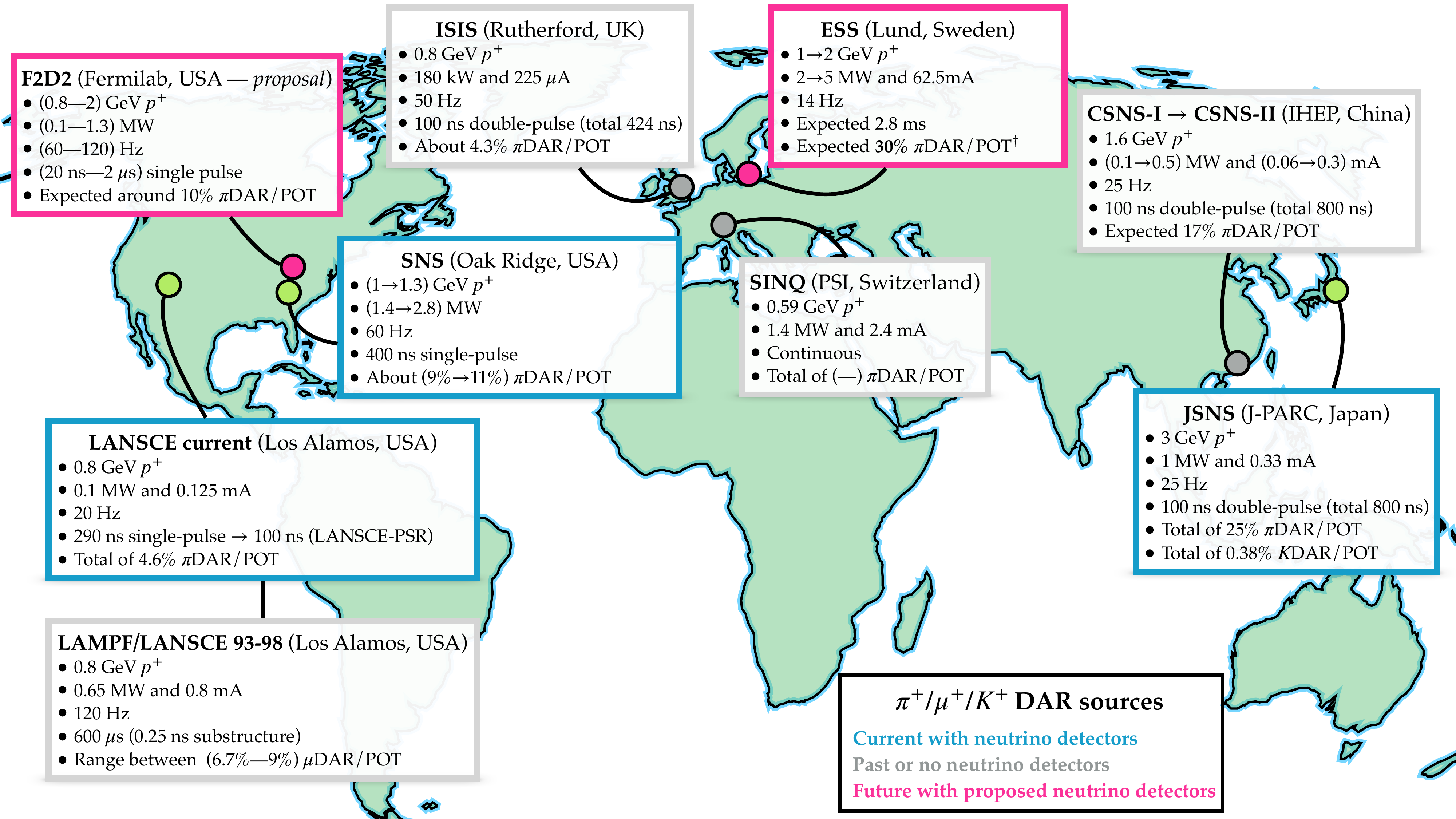}
    \caption{A world map showing the most intense neutron spallation around the world and their parameters of operation.
    \textcolor{NavyBlue}{Current sources} include: the LANSCE spallation source~\cite{CCM:2021leg,CCM:2021jmk,VandeWater:2022qot}, the SNS~\cite{Barbeau:2021exu,Akimov:2022oyb}, the SINQ~\cite{Seidel:2010zz}, the JSNS~\cite{JSNS2:2013jdh,Hino:2021uwz}, and ISIS~\cite{Zeitnitz:1994kz,Burman:1996gt,Thomason:2019fwe}. 
    \textcolor{brightpink}{Future sources} include: the ESS~\cite{Garoby:2017vew,Baxter:2019mcx,Abele:2022iml,Soleti:2023hlr}, the CSNS-II~\cite{2009ChPhC..33....1W,Huang:2015jda,Su:2023klh}, and the proposal for a new accumulator ring for Fermilab's PIP-II~\cite{Pellico:2022dju,Toups:2022yxs}.
    Blue and pink boxes show the current and future sources, respectively.
    We also show the parameters of the LAMPF source during the operation years of the LSND experiment~\cite{LSND:1996jxj,LSND:2001akn}.
    \\
    $^\dagger$The pion yield at ESS is quoted from Ref.~\cite{Baxter:2019mcx}.
    \label{fig:world-map}
    }
\end{figure*}

Spallation sources are most well suited for LLPs lighter than the pion or muon, $m_{\rm LLP} \lesssim m_{\pi^+} \sim 140$~MeV. 
However, despite the larger threshold for kaon production, higher-energy spallation sources, like the JSNS, can still produce enough kaons to  be competitive with conventional accelerator neutrino experiments (see \cref{sec:complementarity}).
For comparison, GEANT4 simulations show that $K^+$ production in the JSNS target is about $3.8 \times 10^{-3}\,K^+$/POT~\cite{Hino:2021uwz,JSNS2:2024bzf}.\footnote{Other simulations with MARS15, however, find that the kaon yield is much larger, $1.1\times 10^{-2}\,K^+$/POT~\cite{Axani:2015dha}.}
This is to be compared, for instance, with the much larger yield at the NuMI absorber, namely  $0.12\,K^+$/POT~\cite{MiniBooNE:2018dus}. 
Despite that, accounting for the difference in POTs between the two sources, the total number of kaons at JSNS will be somewhat smaller than or comparable to the NuMI absorber.
A fair comparison of experimental sensitivity, however, will depend more crucially on background levels and the geometrical acceptance.

In this article, we explore the most minimal LLP models where detectors at spallation sources can place meaningful constraints.
To that end, we focus on a few popular examples, namely: HNLs, the Higgs-portal scalar (HPS), a muonphilic scalar particle (dubbed a muon-portal scalar, MPS), and three specific realizations of ALPs.
The new particles are produced in leptonic or semi-leptonic pion and kaon decays, and in some cases in muon decays.
Production in flavor-changing-neutral-current (FCNC) kaon decays is also possible and dominant for the HPS and for its ALP analogue.

We note that LLPs heavier than $m_\pi$ could in principle also be targeted even without kaon decays if the new particles are abundantly produced by primary $p^+$ collisions at the target via, e.g., bremsstrahlung or Drell-Yann production. 
These channels are more relevant in the forward direction of the proton beam as they tend to produce more boosted and forward-peaked LLPs, rather than the isotropic flux from $\pi^+$ or $\mu^+$ DAR. LSND was located at an angle of $12^\circ$ off-axis with respect to the beam, and so could benefit from these modes~\cite{Foroughi-Abari:2020gju}.
A similar point can be made about LLPs produced in $\eta$ and $\pi^0$ decays.
Therefore, we decide to focus on the isotropic channels where the LLP-DIF signal is, in some cases, mono-energetic.
We leave a study of these additional channels to future literature.

Vector particles could also be considered, however, their production in meson decays is not as efficient when the vector couples to conserved currents~\cite{Pospelov:2008zw}, such as in dark photon or other anomaly-free models.
Dark photons (coupled to SM electric charge via kinetic mixing only), for instance, can be copiously produced in the electromagnetic decay of light mesons, such as $\pi^0 \to \gamma X$ and $\eta \to \gamma X$, however, vector bremsstrahlung at electron and proton beam dumps and fixed targets constrain this parameter space much more effectively.
Therefore, we concentrate on the fermion and (pseudo-)scalar cases where spallation sources can be more sensitive and point the reader to, e.g., \cite{Araki:2023xgb} for a dark photon study at ND280 and \cite{Dev:2021qjj,Foroughi-Abari:2024xlj,Blinov:2024pza,Zhou:2024aeu} for leptophilic vectors at proton accelerators.
We present the details of each of these models after our overview of past and existing facilities.

%%%%%%%%%%%%%%%%%%%%%%%%%%%%%%%%%%%%%%%%%%%%
\section{Overview of Existing Facilities}
\label{sec:sources}

\cref{fig:world-map} shows past (grey), current (blue), and future (pink) spallation sources around the world along with our best estimates for their parameters of operation.
For our LLP study, we focus on locations where neutrino detectors are currently in operation, narrowing our scope to the following three sources:
\begin{enumerate}
\item {\bf Japanese Neutron Spallation Source (JSNS)} at J-PARC. 
JSNS is also particularly interesting thanks to its 1~MW beam, the double short-pulses of 100~ns, and its proximity to multiple types of particle detectors such as the J-PARC Sterile Neutrino Search at JSNS (JSNS$^2$) detectors~\cite{JSNS2:2021hyk} and the near detector of T2K, ND280.
With the construction of new near-detectors for the Hyper-Kamiokande~\cite{Hyper-Kamiokande:2018ofw} experiment, the potential for LLP searches at J-PARC is also likely to improve.
We discuss JSNS in more detail in \cref{sec:JSNS}.
\item {\bf Spallation Neutron Source (SNS)} at the Oak Ridge National Laboratory, where the COHERENT suite of neutrino detectors is located~\cite{Barbeau:2021exu}.
SNS is scheduled for an upgrade in the near future, with the goal of increasing the beam power to $2 - 2.8$~MW divided across two separate target stations~\cite{Akimov:2022oyb}.
Thanks to the existing activity in neutrino physics and the potential for new detectors, such as EOS~\cite{Anderson:2022lbb} or PROSPECT-II~\cite{PROSPECT:2021jey}, SNS is an excellent candidate for LLP searches.
We discuss SNS in more detail in \cref{sec:SNS}.
\item {\bf Los Alamos Neutron Science Center (LANSCE)} at the Los Alamos National Laboratory, where the Coherent-Captain Mills (CCM) neutrino detector is located~\cite{VandeWater:2022qot}.
Due to its sheer volume, CCM is a particularly interesting detector to consider.
The LANSCE facility is also undergoing an upgrade to increase the beam power to $1$~MW~\cite{CCM:2021leg,CCM:2021jmk}, which would further boost its sensitivity to LLPs, especially if the beam pulse were to be shortened.
We discuss LANSCE in more detail in \cref{sec:LANSCE}.
\end{enumerate}
Other spallation sources include the ISIS facility at the Rutherford Appleton Laboratory in the UK~\cite{Zeitnitz:1994kz,Burman:1996gt}, the continuous spallation source SINQ at the Paul Scherrer Institute in Switzerland~\cite{Seidel:2010zz}, and the CSNS facility in China~\cite{Huang:2015jda,Su:2023klh}.
These sources, however, are not currently equipped with neutrino detectors that could search for LLPs.
In the future, the European Spallation Source (ESS) in Sweden~\cite{Garoby:2017vew} could also provide high intensity beams.
ESS plans to increase its beam power to a total of 5~MW with a 14~Hz repetition rate and a long pulse width of about 2.5~ms.
For LLP searches to benefit from the higher intensity at the ESS, more shielding and background rejection techniques would be required to mitigate beam-induced backgrounds.
There are already a host of proposals for fundamental physics at this facility~\cite{Abele:2022iml}, which include a cold neutron program, a CEvNS program~\cite{Baxter:2019mcx}, and searches for HNLs~\cite{Soleti:2023hlr}.

There are also proposals for new compressors, accumulators, a new target station, and beam dump experiments at Fermilab as part of the Accelerator Complex Evolution (ACE) plan for PIP-II~\cite{Pellico:2022dju,Toups:2022yxs}.
Such upgrades could bring the timing and bunch structure of spallation sources to a higher energy beam and benefit from the large-volume neutrino detectors at Fermilab.
Future design studies can readily apply the LLP-DIF signals we discuss to their setups.

% \begin{figure*}[t]
%     \centering
%     \includegraphics[width=\textwidth]{Figs/JPARC_Diagram_Draft-6.png}
%     \caption{A schematic of the J-PARC complex and the neutrino detectors considered in this work. The spallation source is shown in cyan color. The relevant neutrino detectors include the T2K near detector ND280, the KOTO detector at the Hadron Experimental Facility, and the JSNS$^2$ tanks inside and just outside of the Material and Life Science Experimental Facility (MLF). 
%     The Rapid Cycling synchrotron that provides the proton beam for JSNS and the Main Ring synchrotron that provides the beam for T2K and hadron experiments are highlighted in dark blue.\su{I think it is better to remove this figure and just keep the other one, it looks nice but they both have very similar information.}
%     \label{fig:JPARCmap}}
% \end{figure*}

\begin{figure*}[t]
    \centering
    \includegraphics[width=0.9\textwidth]{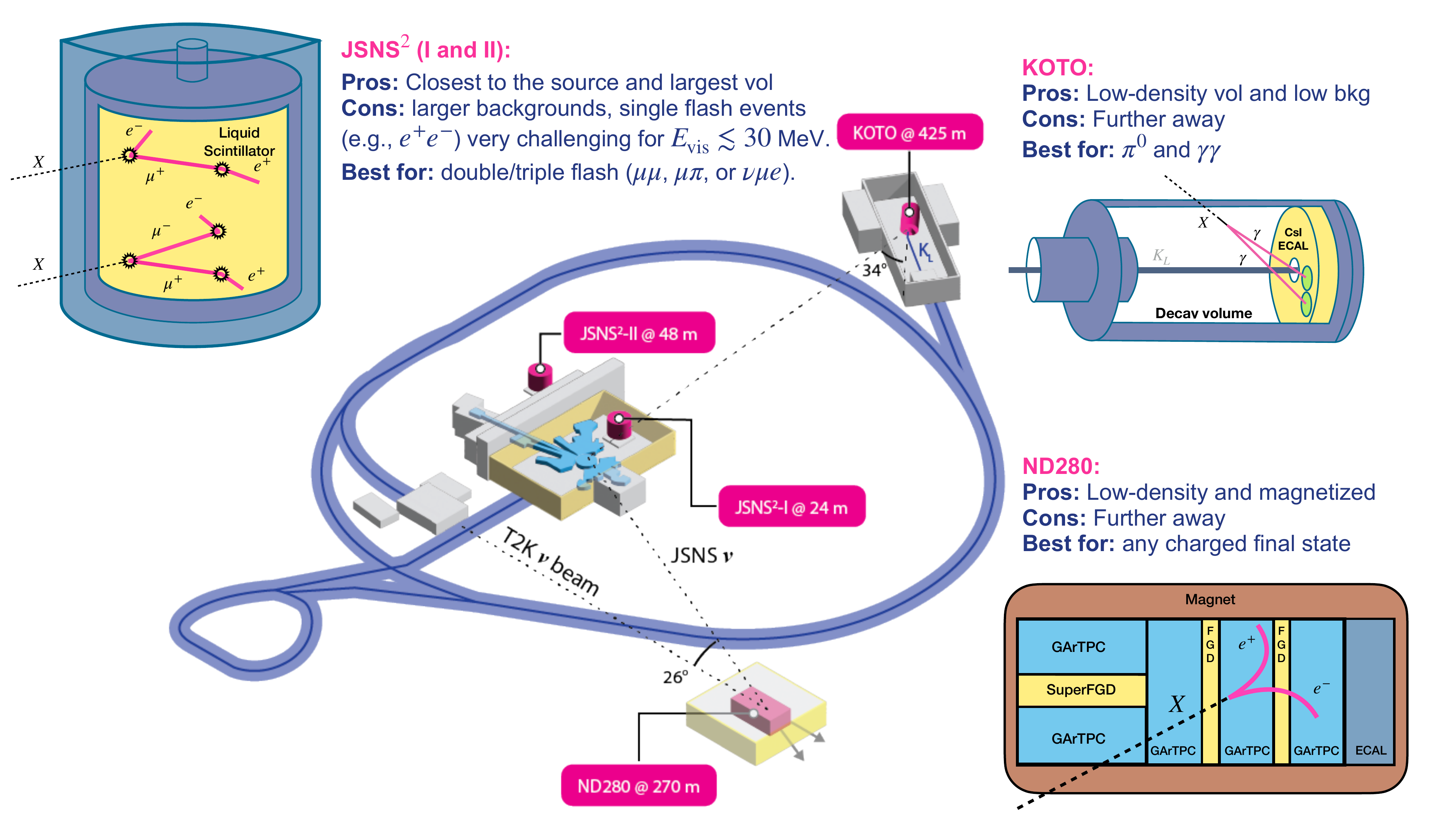}
    \caption{
    A schematic of the J-PARC complex and the neutrino detectors considered in this work. The spallation source is shown in cyan color. 
    The Rapid Cycling synchrotron that provides the proton beam for JSNS and the Main Ring synchrotron that provides the beam for T2K and hadron experiments are highlighted in dark blue.
    The relevant neutrino detectors include the T2K near detector ND280, the KOTO detector at the Hadron Experimental Facility, and the JSNS$^2$ tanks inside and just outside of the Material and Life Science Experimental Facility (MLF). 
    We demonstrate the most promising LLP DIF signatures for each detector according to their detection technology.
    Due to the larger beam energy and kaon decay-at-rest at the JSNS target, these detectors can be used to search for heavier particles produced in, e.g., $K\to \pi X$, which in turn decay to heavier final states such as $X\to \mu^+\mu^-$ and $X\to \pi^+\pi^-$.
    \label{fig:jparc_detectors}}
\end{figure*}

\begin{figure*}[th]
    \centering
    \includegraphics[width=0.49\textwidth]{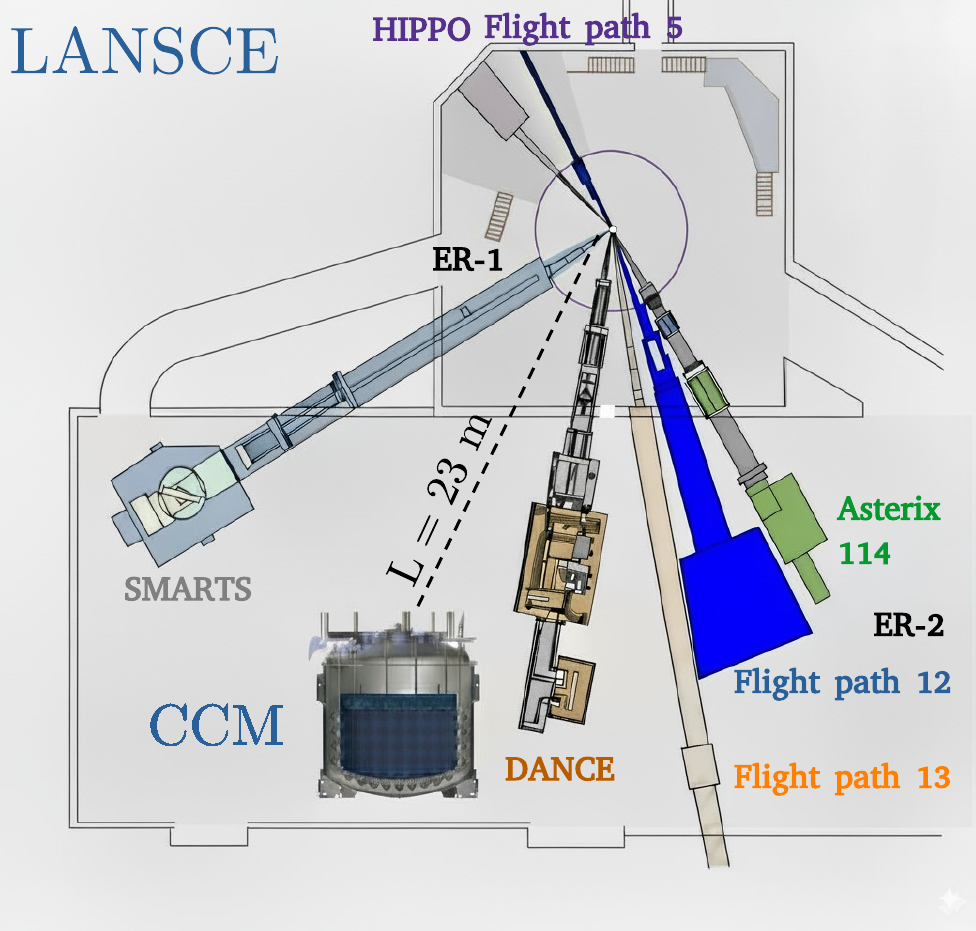}
    \includegraphics[width=0.49\textwidth]{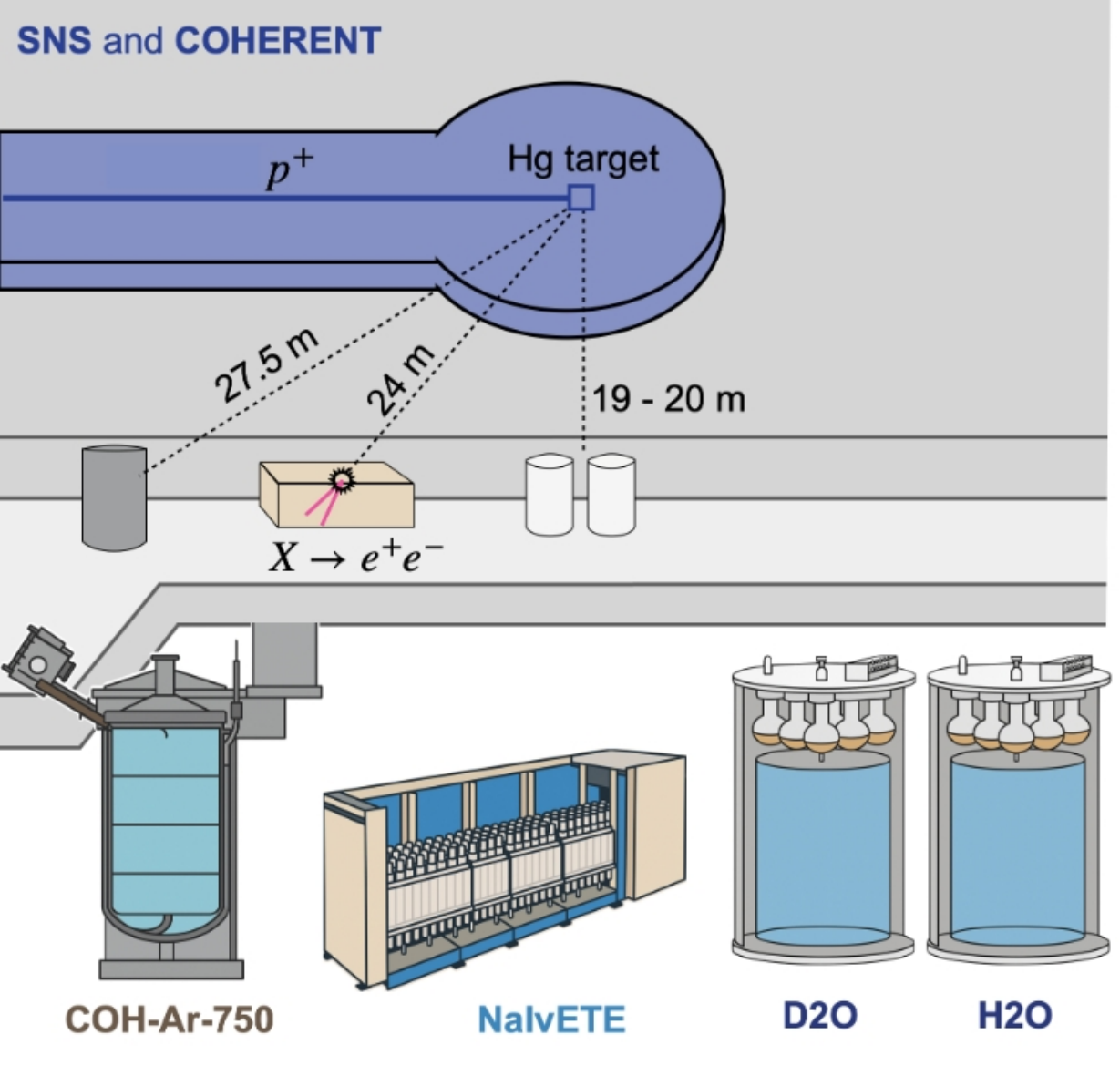}
    \caption{On the left, we show the LANSCE spallation source at the Los Alamos laboratory where the Coherent Captain-Mills detector is located and on the right the Oak Ridge spallation source where the COHERENT detectors are located.
    We highlight the COH-Ar-750, NaIvETe, D2O, and H2O modules are located.}
    \label{fig:LANSCEandOakRidge}
\end{figure*}

The detector technology is also important to consider in deriving the sensitivity to LLPs.
In particular, the energy resolution and directionality can be important leverage to distinguish LLP-DIF from the fast neutrons and neutrinos.
The flux of LLP may be monochromatic when produced in two-body decays of mesons or muons, such as $K \to \pi X_{\rm LLP}$.
In turn, if their decay products are all visible, then they can be identified as a monochromatic signal of charged particle pairs such as $X_{\rm LLP} \to e^+e^-$ or $\gamma\gamma$ final states.

LLPs signals with a continuous energy spectrum can be more challenging to isolate from backgrounds for certain detectors, but could have other distinguishing features such as in the case of $X_{\rm LLP} \to \nu \mu^+ e^-$ decays, where the final state muon decay provides a delayed, second flash of light in the detector.
In addition, the LLP decay can deposit enough energy inside the detector volume to exceed the usual products of neutron or neutrino interactions. Neutron shielding, timing, and particle reconstruction and identification will be crucial for background rejection and a successful new physics program.

%%%%%%%%%%%%%%%%%%%%%%%%%%%%%
\subsection{JSNS}
\label{sec:JSNS}

JSNS is located at J-PARC and benefits from the close proximity to multiple sizable detectors.
The proton beam for JSNS is a $3$~GeV proton beam from the Rapid Cycling Synchroton (RCS) ring with a 1 MW~beam power and a 25 Hz repetition rate.
It has a double pulse structure with each pulse being  $\lesssim 100$~ns in duration and separated by about $\lesssim 600$~ns~\cite{JSNS2:2023hxl}.
The beam impinges on a liquid mercury target to produce about $0.25$ $\pi^+$ and $3.8\times 10^{-3}$ $K^+$ per POT~\cite{JSNS2:2024bzf,JSNS2:2017gzk}.
Out of the sources considered here, this is the only one that can produce kaons in significant quantities.

In addition to JSNS, J-PARC also produces a neutrino beam for the T2K experiment.
The T2K beam is derived from $30$~GeV protons in the Main Ring (MR) synchrotron impinging a thin target.
The pions and kaons are focused by magnetic horns that help collimate the neutrino beam in the direction of the Super-Kamiokande detector, 295~km away.
Neutrinos from the JSNS come with a distinct time profile from the T2K one, offering a high-intensity \emph{pulsed} pion, kaon, and muon DAR flux around the J-PARC complex.

In order to take advantage of the intensity and isotropy of the flux, we consider multiple J-PARC locations:
\begin{itemize}
    \item the multi-component ND280 and its upgrade ND280+ detector at the Neutrino Experimental Facility (NEF), located $280$~m from the T2K target and approximately $270$~m from the JSNS target,
    \item the two liquid-scintillator \JSNSsquared detectors located $24$~m and $48$~m from the JSNS target, respectively. The first detector is located inside the Materials and Life Science Experimental Facility (MLF) and is referred to as the ``inner" detector, while the second one is located outside the MLF and is referred to as the ``outer" detector.
    \item the KOTO photon detector at the Hadron Experimental Facility (HEF), located approximately $425$~m from the JSNS target.    
\end{itemize}
T2K has other near detectors in the same area, such as the NINJA~\cite{NINJA:2020gbg}, WAGASCI-BabyMIND~\cite{BabyMIND:2017mys}, and the INGRID detectors.
However, these are less suited to LLP searches as they are high-density and, with the exception of INGRID, are much smaller detectors.
In the future, there are plans to build an intermediate detector for Hyper-Kamiokande.
One of the proposed detectors is the Intermediate Water Cherenkov Detector (IWCD)~\cite{nuPRISM:2014mzw,Hyper-Kamiokande:2018ofw,HyperKamiokande2025ESPP}, which would be a vertically movable detector located at about $832$~m from the T2K target.
While after its construction, the acceptance of IWCD to LLPs may be the largest at the J-PARC campus, it is still in the design stage and is not expected to be operational until after Hyper-Kamiokande starts. 
For that reason, we leave a detailed study of the IWCD capabilities to future literature.

A sketch of the facilities, detector sites, and the source at JSNS can be found in \cref{fig:jparc_detectors}.
Because each of the detectors above is designed for a different application, they have complementary particle identification capabilities.
This complementarity can improve the experimental coverage of a wide variety of new physics models.

Previous works on LLPs at J-PARC only considered the primary beams for each of the detectors above.
For instance, HNLs in the T2K beam have been constrained by the T2K collaboration using ND280~\cite{Asaka:2012bb,Abe:2019kgx}.
Particles produced at MLF were considered in the context of \JSNSsquared~\cite{Ema:2023tjg}, and those produced in HEF that can be constrained by KOTO were considered in Ref.~\cite{Afik:2023mhj}.
However, to our knowledge, this is the first time that JSNS is considered as a source for ND280(+) or KOTO.

%%%%%%%%%%%%%%%%%%%%%%%%%%%%%%%%%%%%%%%
\subsubsection{The JSNS$^2$ detectors}
\label{sec:JSNS2}

The inner and outer \JSNSsquared detectors contain 17 and 32 tons of liquid scintillator respectively and are designed to search for inverse-beta-decay (IBD) events, $\overline\nu_e p^+ \to e^+ n$, associated with sterile-neutrino oscillations of $\mu$DAR antineutrinos, $\overline\nu_\mu \to \overline\nu_e$.
IBD produces a two-hit signature, where the first hit corresponds to the prompt positron signature and the second-hit to the $\sim 8$~MeV photon emitted by the capture of neutrons on gadolinium.
The \JSNSsquared detectors collect the scintillation light and while some Cherenkov emission can be observed, it is not significant enough to provide an accurate measurement of the direction of the charged particles~\cite{JSNS2:2016zjj}.
Of particular interest to the experiment is the reduction of backgrounds to double-hit events.
Single-hit events are typically associated with fast neutrons produced in the spallation process and are abundant, especially at low energies.
For higher energies, especially above $50$~MeV, the neutron-induced recoil backgrounds quickly decrease.

In our discussion, we will assume that \JSNSsquared cannot distinguish between two different charged particles inside its fiducial volume when these are created at the same location and time.
This means that final states like $e^+e^-$ appear as a single-hit event when the two leptons originate from the same vertex.
On the other hand, muons or pions can lead to cascades of decays (e.g., $\pi\to\mu\to e$) inside the detector with a distinctive timing profile.
In that sense, as long as the pions and muons do not escape the fiducial volume, these particles can be detected via double or triple-pulse coincidences.

\paragraph{Single-hit events ($e^+e^-$):}
The \JSNSsquared collaboration has studied the number of single-hit events in their detector in the energy region and time window of interest for IBD events~\cite{JSNS2:2023hxl}.
Their measurement provides a conservative estimate of the backgrounds of single-hit signatures.

Let us first consider the muon time window, which is defined as events recorded between $2 - 10$~$\mu$s from the beam bunch.
This window targets the neutrinos produced in muon decays at the target.
Considering IBD events, \JSNSsquared investigates this sample in the energy region between $20$~MeV and $60$~MeV.
While not designed to measure the backgrounds to LLP-DIF signatures, the rate in this energy and time window provides an estimate of the background to single-hit signatures of LLPs produced in muon decays, such as HNLs with $m_N < m_\mu - m_e$ decaying to $N \to \nu e^+e^-$.
In this window, the total event rate was found to be $(2.20\pm0.08)\times 10^{-4}/$spill, corresponding to approximately $1.7 \times 10^{5}$ events/year, taking the repetition rate of the beam as $25$~Hz and assuming the detector to be operational $100\%$ time of the year.
Some of this background can be reduced by exploiting the waveform difference between neutron-induced backgrounds and the signal.
Fast-neutron-induced backgrounds arise due to the scintillation light of recoil protons inside the fiducial volume.
This scintillation light has a wider waveform than that of an electron due to the distinct $\dd E/\dd X$ of the two final states.
This is expected to lead to a fast-neutron rejection of $95\%$ at \JSNSsquared for single electron events~\cite{Dodo:2024ulu}.
We expect this number to be similar for $e^+e^-$ final states.
Ultimately, while further experimental efforts may reduce these numbers based on the subtle differences in energy spectrum and reconstruction of the signal and backgrounds, we conclude that it is unlikely that a background-free search can be performed in this channel in the energy region of $20-60$~MeV.

Now we move on to the ``on-bunch" time window, corresponding to signatures associated with pion and kaon decays.
This region is essentially in coincidence with the proton beam bunches and the backgrounds will predominantly come from fast neutrons produced at the target which subsequently produce visible proton recoils inside the detector.
While the collaboration does not provide a number for the total event rate in this time window 0 -- 1.5 $\mu$s, Fig.~1 of Ref.~\cite{JSNS2:2023hxl} indicates that it should be significantly greater than the one quoted for the ``muon" time window due to the larger number of fast neutrons produced at the source.
Therefore, we again conclude that a background-free search for single-hit events in the on-bunch time window is unrealistic. 
Given these difficulties, we still present sensitivity curves for $\text{JSNS}^2$ under the background-free single-hit assumption (shown as dotted lines). 
This is done to illustrate the potential reach in the case where backgrounds could be reduced to a negligible level and to trigger discussion. These signals, however, as we discuss later, can instead be covered by ND280.

\paragraph{Double-hit events ($\mu^\pm e^\mp$):}
When a muon is produced inside the \JSNSsquared detectors, it quickly loses energy and decays to a Michel electron or positron.
This secondary particle then produces a second flash of light, delayed with respect to the initial flash by a few $\mu$s.
This constitutes a clean double-hit event with a characteristic time signature.
Pions produced inside the detector lead to a very similar signature once they promptly decay into a single muon. 
We assume that the decay of the pion cannot be differentiated from its production flash, and so, as far as the timing profile is concerned, pions and muons behave similarly.

The collaboration made a dedicated measurement of Michel electrons induced by fast neutrons by measuring the rate of on-bunch ($t < 2~\mu$s) signals followed by a muon time ($t > 2~\mu$s) Michel electron~\cite{Ajimura:2015yux}. 
This was performed with a small plastic scintillator detector in a few locations inside the building where the first \JSNSsquared detector is located, comparing beam-on and beam-off data (distinct by timing requirements) to estimate the rate of beam-induced fake prompts. 
The signature of Ref.~\cite{Ajimura:2015yux} was $n_{\rm fast} + p(\isotope[12]{C}) \to (\pi^+ \to \mu^+ \to e^+) X$, where $X$ includes all other particles produced in the process.
This was not observed during the dedicated run and an upper limit was set.

We also note that $K^+$DAR neutrinos, already measured by \JSNSsquared~\cite{JSNS2:2024bzf}, also constitute a background to two-hit events through the reaction $\nu_\mu + \isotope[12]C \to \mu^- + \isotope[12]{N}$, albeit in a narrow energy window around $E_{\rm prompt} \lesssim 120$~MeV.

\paragraph{Triple-hit events ($\mu^\pm \mu^\mp$ and $\pi^\pm \mu^\mp$):}  
An even rarer class of signatures includes triple-hit events, where two muon-like particles are produced inside the detector, each producing a subsequent Michel electron.
If a LLP is decaying to such final states, it is already rather heavy, so the energy deposited will be typically much larger than beam-related neutron backgrounds.
As we will see, in many scenarios, this triple-hit signature, under our optimistic assumption of no backgrounds, provides \JSNSsquared a significant advantage over existing searches.

Finally, we note that \JSNSsquared has performed a measurement of $\nu_e + \isotope[12]{C} \to e^- + N_{\rm g.s.}$ events by looking for single electrons followed by the delayed beta decay $\isotope[12]{N} \to e^+ + \nu_e + \isotope[12]{C}$~\cite{JSNS2:2024uxo}.
Because the beta decay is delayed by about $10$~$\mu$s, this signature cannot be used to set limits on the LLPs considered here.

%%%%%%%%%%%%%%%%%%%%%%%%%%%%%%%%%%%%%%%%%%
\subsubsection{The ND280 detector}
\label{sec:ND280}

The T2K near detector, ND280, is an intricately partitioned and magnetized detector located $280$~m from the T2K target and at a slight off-axis angle of $2.04^\circ$. 
Recently, the detector has been upgraded~\cite{T2K:2019bbb}, so we will distinguish between the old and new detectors by name, where ND280 is the old detector and ND280+ is the new detector.

The most upstream module of ND280 is the P$0$D detector, a composite of high-Z materials like lead and brass, plastic scintillators, and water bags, arranged in layers. 
It functions as both an electromagnetic calorimeter (ECAL) and an active water target, with a specific design focus on the study of $\pi^0$ production and neutrino cross sections in water. 
The first and third ECALs are solely made up of lead and scintillator plate layers.
The tracking modules, located downstream, consist of three gaseous argon time projection chambers (GArTPC), interspersed with fine-grained scintillator detectors (FGD). 
These components interact with fewer neutrinos and offer a more conducive environment for LLP searches. 
The entire detection device is situated within a 0.2 Tesla horizontal dipole magnetic field and is encased by additional side and back ECALs, along with side muon detectors.
In the upgrade, ND280+, the POD module has been replaced by a new FGD module (SuperFGD) sandwiched in the vertical direction between two new high-angle GArTPCs (HA-TPCs).
These three components are then enclosed in a Time-of-Flight detector that can be used to time the origin of tracks inside ND280+.
This upgrade increases the acceptance of the detector to large-angle muons produced in neutrino interactions.

The most relevant components for decay-in-flight searches are the GArTPCs. 
The fiducial volume of each TPC consisted of $1.7\times1.96\times0.56$~m of pure Gaseous argon~\cite{Lamoureux:2018owo}, which represents $62\%$ of the total TPC volume of $8.96$~m$^3$~\cite{Assylbekov:2011sh,Gilje:2014cwd,Lamoureux:2018owo}.
For ND280+, there are two additional TPCs where the POD modules used to be. 
These have a volume of $1.87\times0.82\times 2.0$~m$^3$ each~\cite{T2K:2019bbb}.
Assuming the same fiducialization fraction for these, results in a total fiducial GArTPC volume of $11.7$~m$^3$ in ND280+.

The detector is designed to measure neutrino interactions from the T2K beam; however, we emphasize that it is also exposed to a substantial flux of JSNS neutrinos and potentially LLPs. 
The line of sight between JSNS and ND280(+) forms an approximate angle of $26^\circ$ relative to that between the T2K target and ND280(+). 
The ND280(+) detector is situated approximately $270$~m from JSNS.
% {The ND280(+) detector is situated approximately $270$~m from the JSNS source, resulting in an acceptance of roughly $2 \times 10^{-5}$ for all LLPs generated in an isotropic decay like the decays at rest.}
In principle, the JSNS and T2K beams are not synchronized and are sharply pulsed, meaning they do not overlap. 
However, to achieve the sensitivity we present, the off-T2K-beam trigger at ND280(+) would have to be capable of triggering on the time-window of the events originating from JSNS.

%%%%%%%%%%%%%%%%%%%%%%%%%%%%%%%%%%%%%%%%%%
\subsubsection{The KOTO detector}
\label{sec:KOTO}

The KOTO detector is designed to measure $K_L \to \pi^0 \nu \overline{\nu}$ from the pencil $K_L$ beam produced by the $30$~GeV MR proton beam impinging on a target at the Hadron Facility Building (HFB)~\cite{KOTO:2018dsc,KOTO:2020prk}.
The detector is composed of an empty cylindrical vessel with an electromagnetic calorimeter on its downstream wall to capture the two photons from pion decay.
LLPs may also infiltrate the vessel and decay to two photons. 
Provided the photons are captured by the calorimeter downstream, they would leave a similar signature to the $K_L$ one with the difference that the decay may originate away from the $K_L$ beam.
This was explored in Ref.~\cite{Afik:2023mhj} to study the KOTO sensitivity to ALPs produced at the HFB target which subsequently decay to two photons, $a \to \gamma \gamma$.
Here, we ask a similar question, focusing instead on the flux of LLPs from JSNS.
In particular, we estimate the event rate for the HNL benchmark model, targeting the $N \to \nu \pi^0$ decays.

Using Google Maps, we estimate the distance between the JSNS source and the detector to be about $\sim 425$~m.
The angle between the length of the KOTO detector (the direction of the $K_L$ beam) and the $30$~GeV proton beam at the HF building is $16^\circ$~\cite{Shimogawa:2010zz}.
We also estimate the angle between the MR beam and the line of sight from JSNS to KOTO to be about $45^\circ$, which results in an angle between the JSNS line of sight and the length of KOTO to be about $34^\circ$.
This is the typical angle that photons from $\pi^0$s produced in HNL decays or from $a \to \gamma \gamma$ make with the surface normal of the CsI calorimeter.

We can estimate the fiducial volume of KOTO's vessel by assuming all photons travel in the same direction as the LLPs, and calculate the volume in which the direction of travel of photons intersects with the CsI calorimeter.
This fiducial volume would contain a cylinder bisected by an angle of $\theta = 34^\circ$ and shifted with respect to the calorimeter layer by a distance $d$.
The gap $d$ ensures that two photons produced at the base of this cylinder develop enough of a spatial separation at the location of the calorimeter.
The radius of the fiducial bisected cylinder is given by $r = (h - d \tan\theta)/2$.
Taking $d = 20$~cm, we find $V = \frac{\pi r^2 l}{2} \simeq 13$~m$^3$, where $l = 2r /\tan\theta$ .
This volume, while only a fraction of the KOTO detector volume is still significant and even larger than the fiducial volume of the GArTPCs in ND280.
However, the significantly larger distance to JSNS makes KOTO typically less sensitive than the other options considered here.

Because the kaon beam in KOTO is derived from the MR proton beam, kaon-beam-induced events will not overlap with the signal from JSNS, which has a different timing structure and is sharply pulsed.
Therefore, as long as events can be recorded in this JSNS-beam-time window, KOTO should expect no background events.
The timing resolution of the CsI detector is about $1$~ns, allowing for strong discrimination between LLPs produced by JSNS pulses ($25$~Hz) and the neutral particles produced at the HFB target ($1/3.75$ Hz)~\cite{Shimogawa:2010zz}.

%%%%%%%%%%%%%%%%%%%%%%%%%%%%%%%%
\subsection{SNS}
\label{sec:SNS}

SNS at Oak Ridge National Laboratory operates a 1.4 MW proton beam that produces neutrons by directing 1 GeV protons onto a thick liquid-mercury target. 
The beam repetition rate is of 60 Hz, producing pulses of protons with a duration of approximately 695 ns.
Alongside neutrons, copious charged pions are generated, with approximately 0.09 $\pi^+$ per POT~\cite{COHERENT:2021yvp}.

%%%%%%%%%%%%%%%%%%%%%%%%%%%%%%%%
\subsubsection{COHERENT suite of detectors}
\label{sec:COHERENT}

The SNS serves as a neutrino source for the COHERENT experiment: a suite of detectors dedicated to studying CEvNS.
After the first observation of CEvNS in 2017 with a small cesium-iodide (CsI) detector~\cite{COHERENT:2017ipa}, the collaboration has deployed a series of additional modules to detect SNS neutrinos on a variety of targets.
These include the CENNS-10 detector~\cite{COHERENT:2020iec}, which detected CEvNS on argon for the first time and confirmed the scaling of the cross section with the square of the number of neutrons, $N^2$. 
Most recently, COHERENT reported the first measurement of CEvNS on germanium using the Ge-mini detector~\cite{COHERENT:2024axu}. 
These measurements of CEvNS have not only further validated the SM but also provided sensitivity to potential new physics signals, such as non-standard neutrino interactions (NSI)~\cite{DeRomeri:2022twg,Rossi:2023brv,Breso-Pla:2023tnz,Coloma:2023ixt,Coloma:2022avw,Abdullah:2022zue,Papoulias:2017qdn,COHERENT:2017ipa,Coloma:2017ncl}, sterile neutrinos~\cite{Papoulias:2017qdn}, neutrino magnetic moments~\cite{Coloma:2022avw,Abdullah:2022zue,Papoulias:2017qdn}, and light mediators~\cite{Coloma:2022avw,Abdullah:2022zue,Papoulias:2017qdn}, among others.

CEvNS, however, is not the only type of measurement being performed. 
Some detector modules are also well-suited for measurements above the MeV energy scale, as would be required for an LLP search.
For the remainder of this section, we expand on these, focusing on the water modules, NaIvETe, and the LAr detector.
While not an exhaustive list, it provides three examples of the untapped potential of SNS to search for LLPs.
We note that future detectors could also be considered and would benefit from the beam upgrade parameters and the addition of a second target station~\cite{Asaadi:2022ojm}.

%%%%%%%%%%%%%%%%%%%%%%%%%%%%%%%%%%%
\textbf{D2O and H2O:}
The COHERENT collaboration has deployed a heavy water (D$_2$O) detector to be eventually paired with a water (H$_2$O) detector to extract the neutrino flux at SNS using the well-known process $\nu_e+d \to e^+ +n+n$~\cite{COHERENT:2021xhx}. 
These detectors will be placed approximately 19 to 20~m from the SNS target and are referred to as the ``water modules" here.
Each module will contain $592$~kg of target water and will be equipped with a set of PMTs to collect Cherenkov light produced by neutrino interactions in its main volume.
In the H$_2$O module, the collaboration can directly measure $\nu_e + ^{16}O$ CC events in the absence of $\nu_e+p$ or $\nu_e+d$ CC interactions to control this background in the D$_2$O module.
In three years of data taking, the collaboration expects to be able to constrain the neutrino flux to better than $3\%$ using $\nu_e+d$ CC break-up reactions.
As detectors designed for MeV-scale physics, the water modules would be well suited for the signatures discussed here, especially for high-energy and mono-energetic LLP-DIF signals, as these could easily dominate over neutrino scattering events.

%%%%%%%%%%%%%%%%%%%%%%%%%%%%%%%%%%%
\textbf{NaIvETe:}
The COHERENT collaboration is also updating the NaIvE detector to the larger NaI Neutrino Experiment TonnE-scale (NaIvETe) detector~\cite{COHERENT:2023ffx}. 
NaIvETe is designed to utilize sodium iodide (NaI[Tl]) crystals to achieve more precise measurements of CEvNS interactions on the relatively light 
\isotope[23]{Na} nucleus. 
The detector features a modular design, consisting of multiple modules each containing $63$~NaI[Tl] crystals.
Here we assumed that NaIvETe will eventually comprise 7 such modules with a total mass of $7\times (63 \times 7.7$~kg)$\simeq 3.4$~tonnes or approximately $0.92$~m$^3$ in volume assuming a constant crystal density of $3.68$~g/cm$^{3}$. 
A key feature of NaIvETe is its dual-gain PMT bases, which enable the simultaneous readout of keV and MeV scale energy depositions, thereby enabling the LLP-DIF search discussed here. 
NaIvETe can therefore simultaneously pursue a measurement of CEvNS on \isotope[23]{Na}, $\nu_e$ CC interactions on \isotope[127]{I}~\cite{Major:2024wek}, as well as search for the decay in flight of the LLPs discussed here.

%%%%%%%%%%%%%%%%%%%%%%%%%%%%%%%%%%%
\textbf{LAr-750:}
The COHERENT collaboration is also set to deploy the LAr-750 detector, a 750 kg ($610$~kg fiducial) single-phase LAr calorimeter~\cite{tayloe2019cenns750}.
LAr-750 will achieve a recoil energy threshold of approximately $20$~keV in nuclear recoils, building on the operation of the CENNS-10 detector~\cite{COHERENT:2019iyj}. 
The scintillation light produced by the argon target within its active volume will be detected by an array of PMTs, enabling both a precision measurement of CEvNS cross-section on argon but also measurements of CC $\nu_e-\isotope[40]{Ar}$ reactions.
The detector will also demonstrate the feasibility of a future multi-ton scale design that could further boost the sensitivity to LLPs discussed here~\cite{Asaadi:2022ojm}.

\renewcommand{\arraystretch}{1.5}
\begin{table*}[th]
\centering
\footnotesize
\begin{tabular}{|
  >{\centering\arraybackslash}p{1.8cm} |
  >{\centering\arraybackslash}p{2.1cm} |
  >{\centering\arraybackslash}p{1.6cm} |
  >{\centering\arraybackslash}p{3.5cm} |
  c |
  >{\centering\arraybackslash}p{2.2cm} |
  >{\centering\arraybackslash}p{1.3cm} |
  >{\centering\arraybackslash}p{2.5cm}|
}
\hline
\hline
{Source} & {Detector} & {Fiducial Volume} & {Integrated POT} & {Distance} & {Efficiency} & $N_{\rm sig}$ & {Technology} \\
\hline
\hline
\multirow{3}{*}{SNS} & D2O+H2O & $1.08$~m$^3$ & $4.5 \times 10^{23}$ (3 years) & 19.5~m & \multirow{3}{*}{$20\%$} & \multirow{3}{*}{2.44} & D$_2$O + H$_2$O modules \\\cline{2-3}\cline{5-5}\cline{8-8}
& NaIvETe & $0.92$~m$^3$ & ($9.0\%\,\pi^+_{\rm DAR}$/POT) & 24~m & & & NaI[Tl] crystal scintillator \\\cline{2-3}\cline{5-5}\cline{8-8}
& LAr-750 & $0.47$~m$^3$ & & 27.5~m & & & Liquid argon (scintillation) \\
\hline\hline
\multirow{3}{*}{JSNS} & JSNS$^2$-I and II & $20.1$ and $36.6$~m$^3$ & $1.1\times10^{23}$ (3 years) & 24 and 48~m & $20\%$ ($e^+e^-$) \newline ($70\%$ otherwise) & 2.44 & Liquid scintillator (scintillation) \\\cline{2-3}\cline{5-8}
& ND280 & $11.7$~m$^3$ & ($25\%\,\pi^+_{\rm DAR}$/POT) & 270~m & $20\%$ & 2.44 & Gaseous argon \newline (TPC) \\\cline{2-3}\cline{5-8}
& KOTO & $\sim 13$~m$^3$ &($0.38\%\,K^+_{\rm DAR}$/POT) & 425~m & $20\%$ & 2.44 & Vacuum and CsI calorimeter wall\\\cline{2-8}
\hline\hline
LANSCE & CCM & 5~m$^3$ & $2.25\times10^{22}$ \newline(3 years) \newline($4.6\%\,\pi^+_{\rm DAR}$/POT) & 23~m & $20\%$ & 2.44 & Liquid argon (scintillation and Cherenkov) \\\cline{2-8}
\hline\hline
Lujan (deactivated) & LSND (deactivated) & 138~m$^3$ & { $1.69\times 10^{23}$ \newline(1994-1998) \newline(average $7.7\%\,\pi^+_{\rm DAR}$/POT) } & $29.8$~m &  \cref{app:LSND} & 55 & Liquid scintillator (scintillation and Cherenkov)\\\cline{2-8}
\hline\hline
ISIS & KARMEN (deactivated) & $32.7$~m$^3$ & $4.65\times10^{22}$ \newline ($4.3\%$ $\pi^+_{\rm DAR}$/POT) & $17.7$~m & \cref{app:KARMEN} & 4.99  & Liquid scintillator (scintillation) \\\cline{2-8}
\hline 
\hline
\end{tabular}
\caption{Summary of neutrino detector and particle detectors at various spallation sources. The fiducial volume is the amount of active material in the detector that is assumed for the LLP search. The integrated POT is the total number of protons on target that the detector has been exposed to or is expected to be exposed to in the future. The distance is the estimated distance from the target to the (center) detector. The technology refers to the type of detector. The Lujan source is currently deactivated and the LSND and KARMEN detectors are no longer in operation. 
In addition to the signal efficiencies shown, we require $E_{\rm vis}>5$~MeV in all of our sensitivity studies. 
Limits or sensitivities are shown requiring a total number of signal events greater than $N_{\rm sig}$. 
\label{tab:spallation_detectors}
}
\end{table*}

\newcommand{\nobump}{\phantom{\checkmark}}

\newcommand{\flash}{\checkmark} % (you already had this)

\renewcommand{\arraystretch}{1.5}
\begin{table*}[t]
\centering
\begin{tabular}{|c|c|c|c|>{\centering\arraybackslash}p{2.8cm}|>{\centering\arraybackslash}p{2.8cm}|}
\hline\hline
Model & Production & Decay & Timing signature & Mono-energetic Signal & Detectors \\
\hline\hline
\multirow{5}{*}{Heavy Neutral Leptons}
  & $\mu^+ \to e^+\nu N$                     & $N \to \nu e^+e^-$                         &  \begin{minipage}{.125\textwidth}
        \vspace{1ex}
        \includegraphics[width=\textwidth]{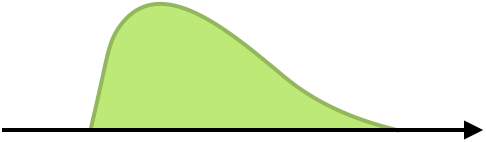}\\[-0.5ex]
    \end{minipage}     & \nobump & \JSNSsquared, ND280, CCM and COHERENT \\
\cline{2-6}
  & \multirow{4}{*}{$\pi^+/K^+ \to \ell N$}  & $N \to \nu e^+e^-$                         & \begin{minipage}{.125\textwidth}
        \vspace{1ex}
        \includegraphics[width=\textwidth]{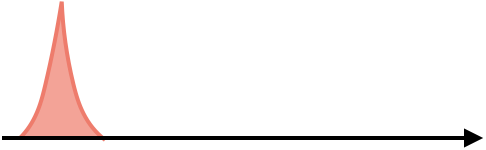}\\[-0.5ex]
    \end{minipage}    & \nobump & \JSNSsquared, ND280, CCM and COHERENT \\
\cline{3-6}
  &                                            & $N \to \nu \mu^+e^- / \pi^+e^-$            & \begin{minipage}{.125\textwidth}
        \vspace{1ex}
        \includegraphics[width=\textwidth]{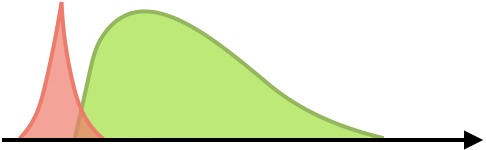}\\[-0.5ex]
    \end{minipage}  & \nobump\!$^{\dagger}$ & \JSNSsquared and ND280 \\
\cline{3-6}
  &                                            & $N \to \nu \mu^+\mu^- / \pi^+\mu^-$        & \begin{minipage}{.125\textwidth}
        \vspace{1ex}
        \includegraphics[width=\textwidth]{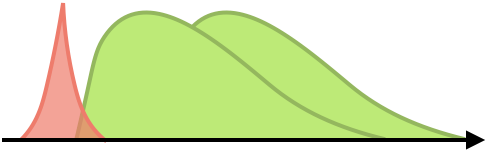}\\[-0.5ex]
    \end{minipage}  & \nobump\!$^{\dagger}$ & \JSNSsquared and ND280 \\
\cline{3-6}
  &                                            & $N \to \nu \pi^0$                          & \begin{minipage}{.125\textwidth}
        \vspace{1ex}
        \includegraphics[width=\textwidth]{Figs/no_delayed.png}\\[-0.5ex]
    \end{minipage}    & \nobump   & KOTO \\
\hline\hline
\multirow{3}{*}{Higgs Portal Scalar}
  & \multirow{3}{*}{$K^+ \to \pi^+ S$}        & $S \to e^+e^-$                             & \begin{minipage}{.125\textwidth}
        \vspace{1ex}
        \includegraphics[width=\textwidth]{Figs/no_delayed.png}\\[-0.5ex]
    \end{minipage}    & \flash & \JSNSsquared and ND280 \\
\cline{3-6}
  &                                            & $S \to \mu^+\mu^- / \pi^+\pi^-$            & \begin{minipage}{.125\textwidth}
        \vspace{1ex}
        \includegraphics[width=\textwidth]{Figs/double_pulse.png}\\[-0.5ex]
    \end{minipage}  & \flash & \JSNSsquared and ND280 \\
\cline{3-6}
  &                                            & $S \to \pi^0\pi^0$                          & \begin{minipage}{.125\textwidth}
        \vspace{1ex}
        \includegraphics[width=\textwidth]{Figs/no_delayed.png}\\[-0.5ex]
    \end{minipage}    & \flash & KOTO \\
\hline\hline
\multirow{2}{*}{Muon Portal Scalar}
  & $\mu^+ \to e^+\nu\nu S_M$                 & $S_M \to \gamma\gamma$                     & \begin{minipage}{.125\textwidth}
        \vspace{1ex}
        \includegraphics[width=\textwidth]{Figs/delayed.png}\\[-0.5ex]
    \end{minipage}     & \nobump &  KOTO, \JSNSsquared, ND280, CCM and COHERENT \\
\cline{2-6}
  & $\pi^+/K^+ \to \mu^+\nu_\mu S_M$          & $S_M \to \gamma\gamma$                     & \begin{minipage}{.125\textwidth}
        \vspace{1ex}
        \includegraphics[width=\textwidth]{Figs/no_delayed.png}\\[-0.5ex]
    \end{minipage}    & \nobump &  KOTO, \JSNSsquared, ND280, CCM and COHERENT \\
\hline\hline
\multirow{2}{*}{ALP: Higgs Coupling}
  & \multirow{2}{*}{$K^+ \to \pi^+ a_{h}$}    & $a_{h} \to e^+e^-$                         & \begin{minipage}{.125\textwidth}
        \vspace{1ex}
        \includegraphics[width=\textwidth]{Figs/no_delayed.png}\\[-0.5ex]
    \end{minipage}    & \flash &  \JSNSsquared and ND280 \\
\cline{3-6}
  &                                            & $a_{h} \to \mu^+\mu^-$                     & \begin{minipage}{.125\textwidth}
        \vspace{1ex}
        \includegraphics[width=\textwidth]{Figs/double_pulse.png}\\[-0.5ex]
    \end{minipage}  & \flash &  \JSNSsquared and ND280 \\
\hline\hline
\multirow{1}{*}{ALP: Flavor Violating}
  & $\mu^+ \to e^+ a_\ell\;(\gamma)$          & $a_\ell \to e^+ e^-$                       & \begin{minipage}{.125\textwidth}
        \vspace{1ex}
        \includegraphics[width=\textwidth]{Figs/delayed.png}\\[-0.5ex]
    \end{minipage}     & \flash &  \JSNSsquared, ND280,  CCM and COHERENT \\
\hline\hline
\multirow{2}{*}{ALP: Flavor Conserving}
  & $\mu^+ \to e^+\nu\nu a_\ell$              & $a_\ell \to e^+ e^-$                       & \begin{minipage}{.125\textwidth}
        \vspace{1ex}
        \includegraphics[width=\textwidth]{Figs/delayed.png}\\[-0.5ex]
    \end{minipage}     & \nobump &  \JSNSsquared, ND280,  CCM and COHERENT \\
\cline{2-6}
  & $\pi^+/K^+ \to e^+ \nu_e a_\ell$          & $a_\ell \to e^+ e^-$                       & \begin{minipage}{.125\textwidth}
        \vspace{1ex}
        \includegraphics[width=\textwidth]{Figs/no_delayed.png}\\[-0.5ex]
    \end{minipage}    & \nobump &  \JSNSsquared, ND280,  CCM and COHERENT \\
\hline\hline
\end{tabular}
\caption{Summary of production/decay channels and timing. The ``Monoenergetic Signal" column contains a $\flash$ when the signal is a single energy bump (two‑body production at rest followed by a fully visible decay inside the detector). 
$^{\dagger}$For HNL rows listing both $N\to\nu\ell^+\ell^-$ and $N\to\pi^+\ell^-$, only the purely visible two‑body option ($\pi^+\ell^-$) would yield a bump.}
\label{tab:signals}
\end{table*}

%%%%%%%%%%%%%%%%%%%%%%%%%%%%%%%%
\subsection{LANSCE and Lujan}
\label{sec:LANSCE}

The Los Alamos Neutron Science Center (LANSCE) delivers an 800 MeV proton beam with a power of approximately 0.8 MW at a 20 Hz repetition rate. 
It has a single-pulse structure with each pulse being about $280$~ns in duration~\cite{VandeWater:2022qot}.
The beam is directed onto a thick tungsten target at the Lujan Center from above, making a right angle with the experimental hall floor.
The number of charged pions per POT is about 0.057 $\pi^{+}$ per POT, where around $80\%$ decay at rest~\cite{CCM:2021leg}.

In the future, more shielding and a Proton Storage Ring (PSR) upgrade that shortens its duration by about one third could help reduce backgrounds for projects utilizing the LANSCE facility.
Shorter beam pulses and a fast detector are especially advantageous for signals from pion decays which arrive at the detector before most neutrons can.

%%%%%%%%%%%%%%%%%%%%%%%%%%%%%%%%
\subsubsection{Coherent Captain Mill detector}
\label{sec:CCM}

The Coherent Captain-Mills (CCM) experiment utilizes the intense source of LANSCE, located 20~m away. 
The CCM detector features a 10-ton cylindrical LAr cryostat with a fiducial volume of approximately 5 tons. 
In addition to searching for CEvNS, CCM searches for light dark matter produced in the decays in flight of $\pi^0$ at the target through their coherent interaction with argon nuclei in the detector, resulting in recoil energies above $\sim 50\,\mathrm{keV}$~\cite{CCM:2021leg,deNiverville:2012ij}. 
The CCM collaboration has investigated the possibility to search for ALPs produced via mechanisms such as the Primakoff process in electromagnetic showers and their detection through inverse Primakoff scattering or decays to photon pairs.\cite{CCM:2021jmk}.
CCM can also measure $\nu_e$-argon CC cross sections to inform future supernova neutrino detection measurements by the DUNE far detectors~\cite{chavez_estrada_exploring_2024}. 

So far, preliminary results have been obtained from an engineering run conducted in 2019 with the CCM120 detector, which featured 120 photomultiplier tubes. 
This initial run demonstrated sensitivity to unexplored regions of parameter space for LDM and ALPs, despite challenges such as non-optimized shielding, limited beam exposure, and contaminated liquid argon. 
Building on these findings, the upgraded CCM200 detector has been deployed, now equipped with 200 photomultiplier tubes, improved liquid argon purity, enhanced shielding, and a tenfold increase in beam exposure~\cite{Aguilar-Arevalo:2025ues}.
In addition, it was recently demonstrated that the CCM detector is capable of detecting Cherenkov light from signatures of sub-MeV energies~\cite{Aguilar-Arevalo:2025ezm}, a feature that can further reduce backgrounds to the LLP signatures we discuss.

In this work, we consider the upgraded CCM200 detector when evaluating our event-rate sensitivities for LLPs produced in $\pi$DAR and $\mu$DAR.
In the absence of kaon production channels, we can only consider LLP lighter than the pion and the muon.
Therefore, the detector signatures include only $e^+e^-$ and $\gamma\gamma$.

%%%%%%%%%%%%%%%%%%%%%%%%%%%%%%%%%%%%%%%%%%%%%%%%
\subsection{Complementarity with high-energy accelerators}
\label{sec:complementarity}

In addition to low-energy $\pi$DAR/$\mu$DAR sources, LLPs can also be abundantly produced at high-energy accelerator beams. 
The T2K beam, for instance, has already been exploited to set strong limits on HNLs using ND280 and the $K^\pm$ DIF parents~\cite{Asaka:2012bb,Abe:2019kgx}.
Their results have also sparked a number of theoretical follow-ups that extend the derived limits to other new physics scenarios using either pion DIF or neutrino scattering as the source of the new particles~\cite{Arguelles:2021dqn,Arguelles:2022lzs,Araki:2023xgb,Batell:2023mdn,Liu:2024cdi}.

The short-baseline neutrino detectors at Fermilab are another notable example~\cite{MicroBooNE:2015bmn,Machado:2019oxb}.
Three liquid argon Time Projection Chamber (LArTPC) detectors lead the sensitivity to LLPs: MicroBooNE, which is no longer operational, and SBND and ICARUS, which are currently collecting data.
These were respectively located 470\,m, 110\,m, and 600\,m from the target for the 8 GeV Booster beam protons.
The three detectors are on-axis with the Booster Neutrino Beam (BNB), but are also exposed to a flux of neutrinos from the Neutrinos from the Main Injector (NuMI) beam.
The NuMI beam originates from 120 GeV protons in the Main Injector~\cite{Adamson:2015dkw}, which provide an intense source of pions, kaons, and heavy mesons both at the target as well as at the absorber location.
The sensitivity of these detectors to LLPs has been extensively studied in the literature.
This includes several theoretical works on LLPs produced at the NuMI absorber and at the BNB~\cite{Batell:2019nwo,Kelly:2021xbv,Arguelles:2021dqn,Coloma:2022hlv,Berger:2024xqk,Chatterjee:2024duf,Alves:2024feq}, as well as experimental exclusions by MicroBooNE~\cite{MicroBooNE:2019izn,MicroBooNE:2021sov,MicroBooNE:2022ctm,MicroBooNE:2023gmv,MicroBooNE:2023eef} and ICARUS~\cite{ICARUS:2024oqb}.

There are advantages for LLP searches at higher energy experiments. 
For example, higher energy beams come with larger meson production yield (more pions per POT) and are usually coupled with larger detectors.
Ultimately, the meson production yield scales more like the beam power than the beam current, so despite the smaller number of POTs seen at higher energy accelerators when compared to spallation sources, the former can still deliver competitive numbers of mesons to serve as a source of LLPs.
Heavier mesons like $K$, $D$, and $B$ mesons also require substantially larger energies to be produced, opening up the reach to heavier LLPs, especially when these can also be produced by direct bremsstrahlung off of the primary proton collisions.
Finally, the boosted nature of the final products collimate the flux of LLPs towards the detector, which can also enhance the observable flux.

Some disadvantages include the large boost factor of the LLPs. 
As compared to DAR sources, the larger $\gamma$ factors in DIF production suppressed the probability of decay of the LLP inside the detector in the small-coupling (long-lived) regime. 
In addition, neutrino-induced backgrounds can become more severe, especially for single electromagnetic final states. 
Note that DIF production of LLPs also eliminates the opportunity to search for mono-energetic signals, which is unique to 2-body DAR production channels like $K\to \pi X$ or $\pi \to \ell N$.
In the case of a signal, DAR would be an attractive channel through which to learn about the mass of the new particle~\cite{Alves:2024feq}.
So, while it is true that high-energy accelerators can still improve the reach to LLPs in some cases (see the parameter space explored by the even-higher-energy SHiP experiment, for instance~\cite{Albanese:2878604}), spallation sources still hold some advantages for light particles below the kaon mass.

Finally, we note that the Mu2e experiment at Fermilab, which will search for the neutrinoless coherent conversion $\mu^-N \to e^-N$ in aluminium, can also be used for LLP-DIF signals produced from $\mu$DAR~\cite{Mu2e:2014fns}. 
Mu2e will stop $\sim 10^{18}$ muons using 8~GeV protons in 200~ns micro-bunches separated by 1.7~$\mu$s, offering sensitivity to $\mu$-coupled LLPs produced in rare muon decays in a much similar way to the signatures discussed here.

%%%%%%%%%%%%%%%%%%%%%%%%%%%%%%%%%%%
\section{Neutrino Portal ($N$)}
\label{sec:hnls}

\begin{figure*}[t]
    \centering
    \includegraphics[width=\columnwidth]{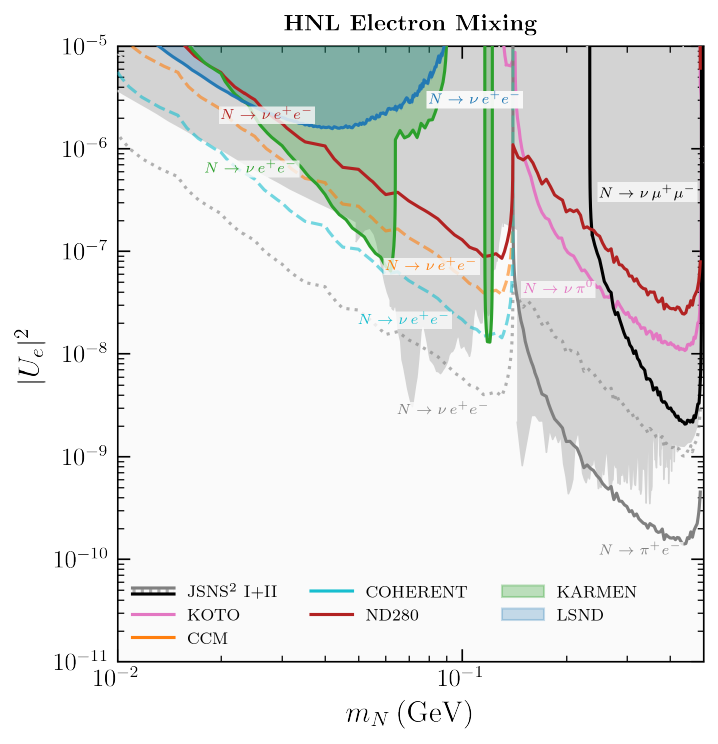}
    \includegraphics[width=\columnwidth]{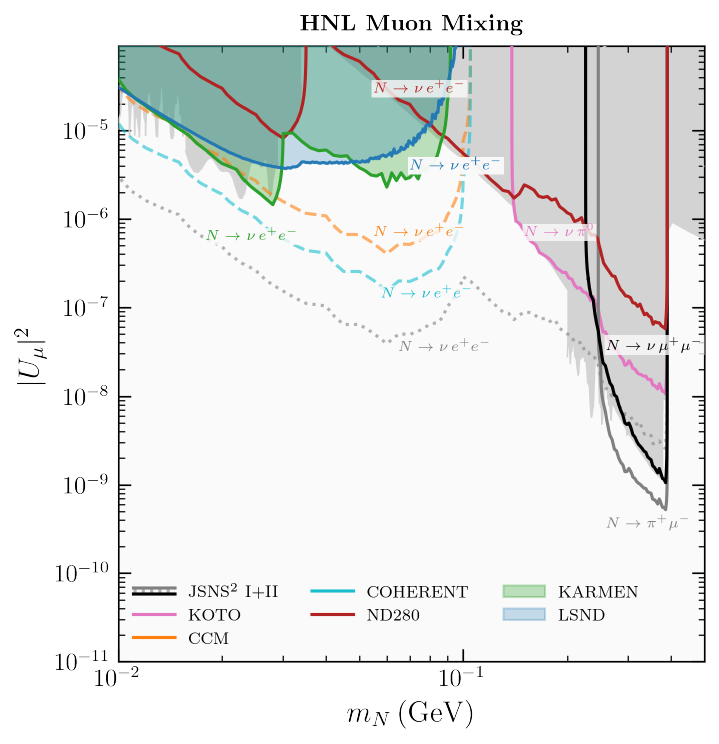}
    \caption{The projected sensitivities to heavy neutral leptons produced in pion, kaon and muon decays. 
    Each color represents a different decay final state, including $N \to \nu e^+ e^-$, $N \to \nu \mu^+ \mu^-$, and $N \to \ell^- \pi^+$. Solid lines indicate the sensitivities expected for the cleanest channels, dashed lines correspond to channels likely to suffer from significant background contamination, and dotted lines denote single-hit signals in \JSNSsquared. The current experimental bounds, extracted from Ref.~\cite{Fernandez-Martinez:2023phj}, are shown in grey.
    We also compute the excluded regions from the past experiments LSND and KARMEN (in shaded), following the procedure outlined in \cref{app:LSND,app:KARMEN}.
    Our LSND lines are in good agreement with Ref.~\cite{Ema:2023buz}. 
    \label{fig:hnl_sensitivity}
    }
\end{figure*}

We begin with the well-motivated extension of the SM by HNLs, singlet fermions that mix with SM neutrinos (see~\cite{Abdullahi:2022jlv} for a review).
This simple addition triggers the seesaw mechanism~\cite{Minkowski:1977sc,Mohapatra:1979ia,GellMann:1980vs,Yanagida:1979as,Lazarides:1980nt,Mohapatra:1980yp,Schechter:1980gr,Cheng:1980qt,Foot:1988aq}, whereby neutrino masses are naturally small due to a hierarchy between the Dirac masses of neutrinos, stemming from the Higgs mechanism, and the Majorana mass of the HNLs.
In low-scale variants of the seesaw~\cite{Mohapatra:1986bd,GonzalezGarcia:1988rw,Wyler:1982dd,Akhmedov:1995ip,Akhmedov:1995vm,Barry:2011wb,Zhang:2011vh}, the mass scale of the HNL can be within reach of accelerator experiments.
In this case, the smallness of neutrino masses is explained by an approximate conservation of lepton number and the HNLs become pseudo-Dirac particles.
Phenomenologically, the HNL behaves like a Dirac particle as long as the mass splitting between its two Weyl components is much smaller than any energy scale in the problem.
We will focus our discussion on this case.

The relevant low-energy interaction Lagrangian is given by
\begin{align}
    -\mathscr{L_{\rm HNL}} & \supset \frac{g}{2 c_W} U_{\alpha N}^* \overline{\nu_\alpha} \slashed{Z} P_L N 
    \\\nonumber
    &\quad+ \frac{g}{\sqrt{2}} U_{\alpha N}^* \overline{\ell}_{\alpha} \slashed{W}  P_L N  + \text{ h.c.},
\end{align}
where the sum over $\alpha \in \{e, \mu, \tau \}$ is implicit. 
For simplicity and following the standard convention in the literature, we assume the flavor structure of the mixing matrix $U$ to have a single flavor dominance.
In particular, we focus on cases where only $U_{e N}$ or $U_{\mu N}$ are non-vanishing.
A non-zero value of $U_{\tau N}$ would not add to the production of HNLs at spallation sources, but it could enhance the decay rate thanks to neutral-current decays, such as $N \to \nu_\tau e^+e^-$ and $N \to \nu_\tau \pi^0$.

The leptonic branching ratio of pions and kaons, $M = \pi, K$, is given by
\begin{equation}\label{eq:KtolN}
    \mathcal{B}(M^+ \to \ell^+_\alpha N) = |U_{\alpha N}|^2 \rho(r_N, r_\ell) \mathcal{B}(M \to \ell^+_\alpha \nu_\alpha),
\end{equation}
where $\rho(x,y) = \lambda^{1/2}(1,x,y) (x+y - (x-y)^2)/(x (1-x)^2)$ and $r_i = m_i/m_M$.
In addition to meson decays, HNLs can also be produced in muon decays, which is particularly relevant for scenarios with dominant $U_{\mu N}$ mixing, as muon decays offer a direct production channel for $N$. 
In the case of $U_{e N}$ mixing, production is also possible but typically less competitive with other constraints in the parameter space.

For $m_N < m_\mu$, the differential decay rates for muon decay into HNLs are given by~\cite{Ema:2023buz}:
\begin{equation}\label{eq:muon_decay_Umu}
\begin{split}
&\frac{d\Gamma\bigl(\mu \to e\,\nu_e\,N\bigr)}{dE_N}
  ={} \frac{G_F^2\,|U_{\mu N}|^2}{12\,\pi^3}
       \sqrt{E_N^2 - m_N^2} \\[-0.3ex]
     {}&\times\bigl[\,3\,E_N\,(m_\mu^2 + m_N^2)
       - 4\,m_\mu\,E_N^2 - 2\,m_\mu\,m_N^2\bigr],
\end{split}
\end{equation}

\begin{equation}\label{eq:muon_decay_Ue}
\begin{split}
\frac{d\Gamma\bigl(\mu \to e\,\nu_\mu\,N\bigr)}{dE_N}
  ={}& \frac{G_F^2\,|U_{e N}|^2}{2\,\pi^3}
       \,E_N\,\sqrt{E_N^2 - m_N^2} \\[-0.3ex]
     {}&\times\bigl[m_\mu^2 + m_N^2 - 2\,m_\mu\,E_N\bigr].
\end{split}
\end{equation}
Once produced, the HNL can decay invisibly into three neutrinos, into two leptons, or hadronically.
The relevant differential decay rates can be found in Ref.~\cite{Coloma:2020lgy}.
In this work, we will focus on the following modes 
\begin{subequations}
\begin{align}
    &N \to \nu e^+e^- \quad & (\text{NC and CC}),
    \\
    &N \to \nu e^\pm\mu^\mp \quad & (\text{CC}),
    \\
    &N \to \nu \mu^+\mu^- \quad & (\text{NC and CC}), 
    \\
    &N \to \nu \pi^0 \quad & (\text{NC}),
    \label{eq:Ntonupi}
    \\
    \label{eq:Ntoepi}
    &N \to e^- \pi^+ \quad & (\text{CC}),
    \\
    &N \to \mu^- \pi^+ \quad & (\text{CC}),
    \label{eq:Ntomupi}
\end{align}
\end{subequations}
and the corresponding charge-conjugate channels for $\overline{N}$.
In the second column, we show the corresponding weak interaction responsible for the decay: neutral-current (NC) or charged-current (CC).
Since the HNLs are (pseudo-)Dirac particles and lepton number is approximately conserved, the charge of the charged lepton produced in \cref{eq:Ntoepi,eq:Ntomupi} is fixed.
The total HNL lifetime is also determined by purely NC decays into neutrinos and into a single photon.
While the latter is a visible signature, it has a smaller decay rate.
We neglect other subdominant modes, like $N \to \nu \pi \pi$ decays.

\subsection{Sensitivities}
Our results for the HNL model in the two single-flavor dominance scenarios are shown in \cref{fig:hnl_sensitivity}. We present both existing constraints and projected sensitivities at future spallation facilities.

The current bounds are derived from LSND and KARMEN data, computed following the procedures in \cref{app:LSND,app:KARMEN}. The use of LSND stopped-muon data was first highlighted in Ref.~\cite{Ema:2023buz}; we reproduce their results in good agreement and extend the analysis by incorporating KARMEN measurements of neutral-current cross sections. We find that KARMEN provides particularly strong limits in the mass range where pion decays dominate HNL production, in some cases yielding the most stringent constraints available in the literature. These bounds are complementary to those from LSND.

For future sensitivities, we consider the JSNS, SNS (COHERENT), and LANSCE (CCM) spallation sources under the assumptions of \cref{sec:sources}. Dashed lines in \cref{fig:hnl_sensitivity} indicate the background-free reach for $e^+e^-$ signatures. While optimistic, these projections highlight the strong motivation for background reduction: if negligible levels could be achieved, substantial improvements over present bounds would be possible, especially for muon mixing.

Among the three facilities, only \JSNSsquared produces a sufficiently intense kaon flux to access multi-hit signatures such as $\mu^+\mu^-$, $\pi^+e^-$, and $\pi^+\mu^-$ (see \cref{sec:JSNS}). For these channels, a background-free analysis is more realistic, and the resulting sensitivities underline the complementarity of different detectors and decay modes. At lower masses, below the muon and pion kinematic thresholds, ND280 offers the most robust reach. 
ND280 benefits from the ability—unlike \JSNSsquared—to cleanly distinguish $e^+e^-$ events. By contrast, at higher masses where kaon decays dominate, \JSNSsquared provides the leading sensitivity. Overall, the future reach is generally weaker or at best comparable to the existing constraints, with the exception of masses near the kaon decay kinematic limit, where modest improvements are expected for both electron and muon mixings.

%%%%%%%%%%%%%%%%%%%%%%%%%%%%%%%%%%%
\section{Scalar Portals}
\label{sec:scalars}

\subsection{Higgs Portal Scalar ($S$)}
\label{sec:hps}

\begin{figure}[t]
    \centering
    \includegraphics[width=\columnwidth]{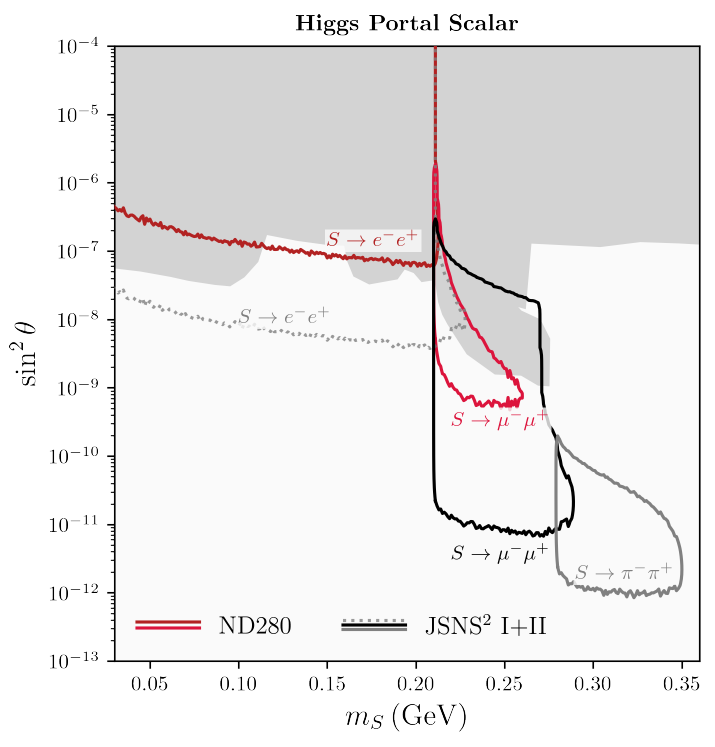}
    \caption{The projected sensitivities to the Higgs portal scalars produced in kaon decays. 
    Each color represents a given decay final state, including $S\to e^+e^-$, $S\to \mu^+\mu^-$, and $S\to \pi^+\pi^-$. Solid lines indicate the sensitivities expected for the cleanest channels and dotted lines denote single-hit signals in \JSNSsquared. The current experimental bounds, extracted from Fig.~131 of Ref.~\cite{Antel:2023hkf}, are shown in grey.
    \label{fig:hps_sensitivity}
    }
\end{figure}

We now move to our second benchmark model, the Higgs portal scalar~(HPS).
This is an extension of the SM by a singlet scalar particle $S$ that mixes with the Higgs boson $H$ through renormalizable operators, $(\mu S + \lambda S^2) H^\dagger H$.
Beyond the model's simplicity, this scenario is also compelling from a theoretical standpoint. 
The scalar $S$ can be identified with the relaxion in cosmological solutions to the Higgs hierarchy problem~\cite{Graham:2015cka,Flacke:2016szy} and could constitute a mediator between the SM and dark matter~\cite{Arcadi:2019lka}.
Phenomenologically, the only ingredients necessary for our analyses are the mass of $S$ and its mixing angle with the SM Higgs.

The low-energy Lagrangian reads
\begin{equation}
\mathscr{L}_S \supset  \frac{1}{2}\partial_\mu S \partial^\mu S - \frac{m_{S}^2}{2}S^2 - \sin{\theta} \, S \sum_{f} \left( \frac{m_f}{v}  \overline{f} f \right),
\end{equation}
where $f$ are the SM fermions that get their mass from the Higgs vacuum expectation value $v$.
The direct couplings of $S$ to the Higgs boson depend on the details of the scalar potential, so we do not discuss them here. 

The scalar is produced primarily in kaon decays via the flavor-changing neutral-current decay $K^+ \to \pi^+ S$, given by~\cite{Leutwyler:1989xj}
\begin{equation}
    \mathcal{B}(K^+ \to \pi^+ S) = \sin^2\theta \frac{ |y_{sd}|^2 m_K^3}{16\pi} \lambda^{1/2}(1, r_S^2, r_\pi^2) \tau_K,
\end{equation}
where $r_i = m_i/m_K$, $\tau_K$ is the kaon lifetime, and
\begin{equation}\label{eq:KtopiS}
    y_{sd} = \frac{3}{32\pi^2 v^3} \sum_{q = u,c,t} V_{qd}^* V_{qs} m_q^2.
\end{equation}
For neutral kaon decays, the rate is instead proportional to $(\Re y_{sd})^2$ for $K_L$ and $(\Im y_{sd})^2$ for $K_S$.
Because the couplings of $S$ to SM fermions are proportional to their masses, the Higgs portal scalar is typically a long-lived particle.
At the masses to which J-PARC experiments are sensitive, the most relevant decay mode of $S$ is into dileptons,
\begin{equation}\label{eq:Stoll}
    \Gamma_{S \to \ell^+\ell^-} = \sin^2\theta \frac{m_\ell^2 m_S}{v^2 8\pi} \left( 1 - \frac{4 m_\ell^2}{m_S^2}\right)^{3/2}.
\end{equation}
For $m_S > 2 m_\pi$, the branching ratios of $S$ are dominated by hadronic modes.
For the range of masses we are interested in ($\leq$ 350 MeV), the hadronic decays are well described by Chiral Perturbation theory and we adopt the rates in Ref.~\cite{Batell:2019nwo},
\begin{equation}
\Gamma_{S \to \pi^-\pi^+}=\sin^2\theta \frac{\left|G_\pi\left(m_S^2\right)\right|^2}{16 \pi v^2 m_S}\left(1-\frac{4 m_\pi^2}{m_S^2}\right)^{1 / 2},
\end{equation}
where $G_\pi(s) = (2 s + m_\pi^2)/9$.
The branching ratio into the charged pions is twice as large as the one into neutral ones.

%%%%%%%%%%%%%%%%%%%%%%%%%%%%%%%%%%%
\subsection{Muon Portal Scalar ($S_M$)}
\label{sec:muonphilic}

\begin{figure*}[t]
    \centering
    \includegraphics[width=0.49\textwidth]{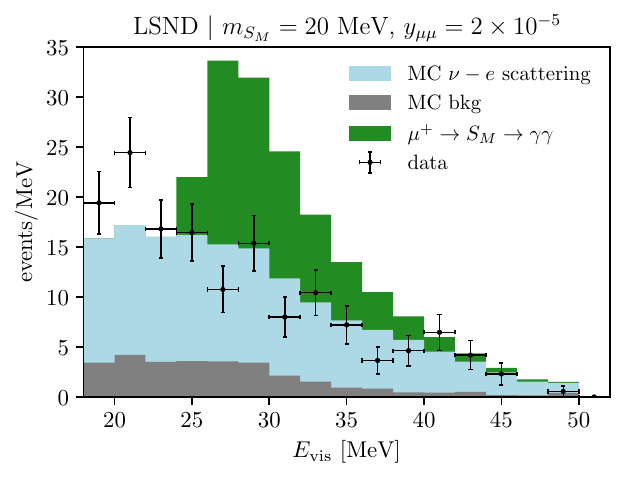}
    \includegraphics[width=0.49\textwidth]{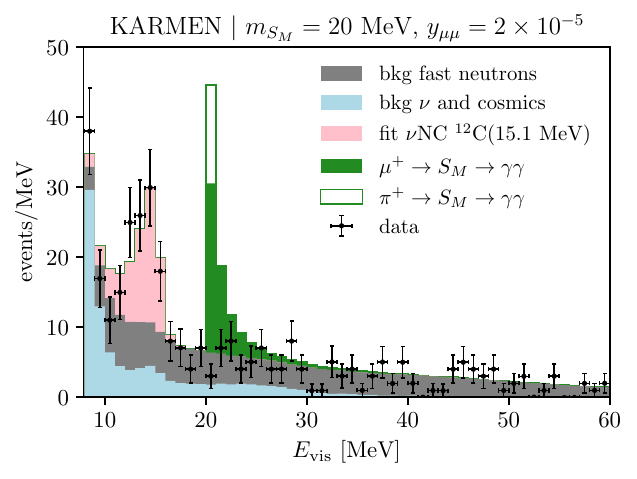}
    \caption{Energy distribution of the events in LSND~\cite{LSND:2001akn} (left) and KARMEN~\cite{KARMEN:1998xmo} (right). 
    We also show a representative signal from the muonphilic scalar scenario in each case for $m_{S_M} = 20$~MeV and $y_{\mu \mu} = 2\times 10^{-5}$.
    For KARMEN, only events and decays inside the neutrino time window are shown (see \cref{eq:neutrino_window}).
    \label{fig:Sm_with_data}
    }
\end{figure*}

\begin{figure}[t]
    \centering
    \includegraphics[width=\columnwidth]{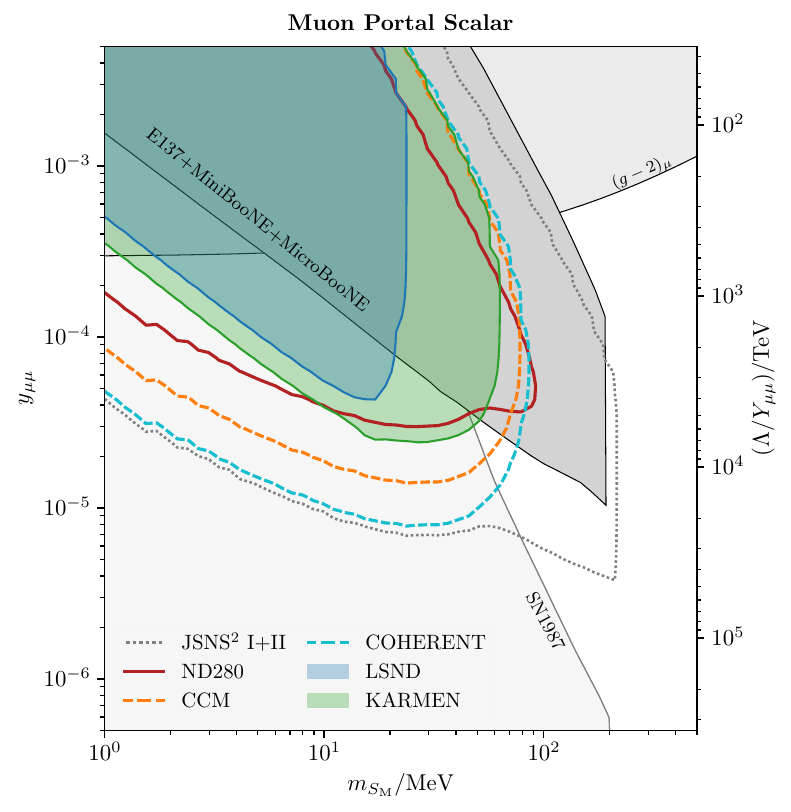}
    \caption{The projected sensitivities of CCM, JSNS$^2$, and KOTO detectors to muonphilic scalars as well as existing limits in grey. Solid lines indicate the sensitivities expected for the cleanest channels, dashed lines correspond to channels likely to suffer from significant background contamination, and dotted lines denote single-hit signals in \JSNSsquared. The E137 limits are taken from~\cite{Dutta:2023fnl,Dobrich:2015jyk}, MicroBooNE and MiniBooNE limits from ~\cite{Cesarotti:2023udo}.
    Supernova limits reinterpreted from~\cite{Croon:2020lrf} are shown as a lighter grey tone. 
    \label{fig:muonphilic_sensitivity}}
\end{figure}

We now turn to the case of a muon-specific force mediated by a scalar particle, the muonphilic scalar $S_M$~\cite{Batell:2016ove}.
The low-energy Lagrangian of the model is
\begin{align}\label{eq:muonphilic}
\mathscr{L}_{\rm S_M} &\supset  \frac{1}{2}\partial_\mu S_M \partial^\mu S_M  
\\\nonumber 
&- \frac{m_{S_M}^2}{2} S_M^2 -  y_{\mu\mu} S_M\overline{\mu}\left(c_S +c_P \gamma_5\right)\mu.
\end{align}
The scalar only couples to muons and is, therefore, less constrained than the other scenarios considered so far~\cite{Ariga:2023fjg,Cesarotti:2023sje,Blinov:2024gcw}.
For simplicity, we will stick to the pure scalar case, $c_S = 1$ and $c_P=0$.
The coupling to muons may originate from a dimension-5 operator such as $\frac{Y_{\mu \mu}}{\Lambda} S_M \overline{L_\mu} H \mu_R$ such that $y_{\mu \mu} = Y_{\mu \mu} v/\sqrt{2}\Lambda$.

At spallation sources, $S_M$ production can occur in both meson and muon decays. 
Pion decays contribute to $S_M$ production only for $m_{S_M} < m_\pi - m_\mu \simeq 30$~MeV via the three-body channel $\pi^+ \to \mu^+ \nu_\mu S_M$, providing a flux of $S_M$ during the on-bunch time window. 
For masses in the range $30~\text{MeV} < m_{S_M} < m_\mu \simeq 105$~MeV, production is dominated by muon decay ($\mu^+ \to e^+ \nu_e \overline{\nu}_\mu S_M$), while for larger masses we rely on kaon decays ($K^+ \to \mu^+ \nu_\mu S_M$) which will only be relevant for JSNS. 
For the $\pi$ and $K$ three-body decays, we use the expressions from Ref.~\cite{Krnjaic:2019rsv}, while for the $\mu$ four-body decay, we have derived the corresponding expression in \cref{app:four-body-leptonic} by generalizing the result of Ref.~\cite{Ema:2025bww}. 
We have simulated the leptonic four-body decays using \textsc{MadGraph5} (v3.5.0)\cite{Alwall:2014hca}, finding excellent agreement with the analytical results. 
The resulting branching ratios are shown in \cref{fig:branchings_Sm}.
%The scalar energy spectrum is interpolated for logarithmically spaced values of $m_{S_M}$ across the full kinematic range, and the total decay rate is rescaled according to the value of $y_{\mu\mu}$. 

When $m_{S_M} < 2 m_\mu$, the scalar is long lived and can only decay via the muon-loop-induced coupling to photons.
More specifically, with $x = 4 m_\mu^2/m_{S_M}^2$, the decay rate into two photons is
\begin{equation}
\Gamma_{S_M \to \gamma\gamma} = \frac{\alpha^2 m_{S_M}^3}{64\pi^3} \left| \frac{y_{\mu \mu}}{m_\mu} x [1 + (1 - x) f(x)] \right|^2,
\end{equation}
with the loop function
\begin{equation}
f(x) = 
\begin{cases}
\arcsin^2(\sqrt{x^{-1}}), & x > 1 \\
-\frac{1}{4} \left[ \ln\left(\frac{1+\sqrt{1-x}}{1-\sqrt{1-x}}\right) - i\pi \right]^2, & x \leq 1
\end{cases}.
\end{equation}

Since we consider production via kaon decays, we can also explore the region $m_{S_M} > 2 m_\mu$. 
For these masses, the decay $S_{M} \to \mu^+ \mu^-$ becomes kinematically accessible. 
However, this additional channel renders the $S_{M}$ very short-lived, and we have verified that the resulting $S_{M}$ flux reaching the \JSNSsquared detector is too small to provide any interesting sensitivity.

We show additional limits from E137~\cite{Dutta:2023fnl,Dobrich:2015jyk} that target the scalar coupling to photons in an electron beam dump as well as limits from the muon $(g-2)$.
With the latest theoretical predictions~\cite{Aliberti:2025beg} including the lattice results for the hadronic vacuum polarization diagrams~\cite{Borsanyi:2020mff,Bazavov:2024eou,Djukanovic:2024cmq,RBC:2024fic} and experimental measurements~\cite{Muong-2:2025xyk}, we require that
\begin{equation}
    \Delta a_\mu^{\rm new-physics} < 165.9 \times 10^{-11},
\end{equation}
at $2\sigma$ based on the difference $\Delta a_\mu = a_\mu^{\rm EXP} - a_\mu^{\rm TH} = (384 \pm 637) \times 10^{-12}$.

Other constraints include the results of Ref.~\cite{Cesarotti:2023udo}, which the use electromagnetic final states in neutrino detectors in the Booster Neutrino Beam, namely MicroBooNE and MiniBooNE, to new limits on $S_M$. 
The NA64 collaboration has also placed limits by looking for missing energy in muon-beam fixed target setup~\cite{NA64:2024nwj}.
We note that these constraints, however, assumed the scalar to be invisible in the experiment, which may not be the case in the large coupling region where it applies, especially when $m_{S_M} > 2m_\mu$, so we omit it.
Other proposals to constrain this model include muon fixed target experiments and future beam dumps~\cite{NA64:2018iqr,Kahn:2018cqs,Forbes:2022bvo,Cesarotti:2022ttv}.
Again, such missing energy and missing momentum techniques are sensitive to models where $S_M$ decays to invisible dark matter particles~\cite{Chen:2018vkr}, while our bounds assume that $S_M$ couples exclusively to muons.

\subsection{Sensitivities}

\Cref{fig:hps_sensitivity} shows the compilation of experimental limits and sensitivities to the HPS. 
Since the HPS is produced via kaon decays, among the three current spallation sources only JSNS is energetic enough to yield a substantial kaon flux. 
We therefore present only the sensitivity of the J-PARC facility, highlighting the complementarity of the different detectors. The results are shown separately for the three visible decay channels: $e^+e^-$, $\mu^+\mu^-$, and $\pi^+\pi^-$. At low masses, decay-in-flight (DIF) searches compete with strong limits from kaon decays $K^+ \to \pi^+ S$, which do not require observation of the scalar decay and scale as $\sin^2\theta$. By contrast, our signal scales as $\sin^4\theta$ in the small coupling (long-lived) regime. 

The ND280 sensitivity improves only midly on existing limits. 
For $\text{JSNS}^2$, we also display the background-free sensitivity line for single-hit $e^+e^-$ events; however, as discussed in~\cref{sec:JSNS2}, the actual backgrounds are expected to be substantial, and this line should therefore be regarded as extremely optimistic. 
Once the scalar becomes sufficiently heavy to decay into muon or pion pairs, it becomes significantly shorter-lived, which strengthens the upper bound and relaxes the lower bound. In this mass region, \JSNSsquared would be able to improve on previous searches by more than an order of magnitude, thanks to the high intensity of JSNS as well as the proximity and large decay volume of the \JSNSsquared detectors.

\cref{fig:Sm_with_data} shows the number of events at LSND and KARMEN as a function of reconstructed visible energy.
An example of a new physics prediction is shown in green for a representative point in the parameter space of the MPS model: $m_{S_M} = 20$~MeV and $y_{\mu \mu} = 2\times 10^{-5}$.
Clearly, this point can be seen to be excluded by the data.
In KARMEN, only events that decay within the neutrino window are shown, including both muon (filled green) and pion (unfilled green) production of MPS.

\cref{fig:muonphilic_sensitivity} shows the experimental limits and sensitivities for the MPS model.
All lines correspond to the decay $S_M \to \gamma\gamma$. 
The LSND and KARMEN constraints (shown shaded) set the strongest laboratory bounds at low masses, exceeding those from MicroBooNE and E137. 
These regions, however, are already excluded by supernova limits, which dominate most of the parameter space at low masses. 
The projected sensitivity of ND280 is expected to remain close to the current bounds. 
At higher masses, where supernova constraints weaken significantly, only low-background $\gamma \gamma$ searches from CCM, COHERENT, and in particular \JSNSsquared could improve upon existing limits.

%%%%%%%%%%%%%%%%%%%%%%%%%%%%%%%%%%%
\section{Axion-Like Particles}
\label{sec:ALPs}

The strong CP problem in Quantum Chromodynamics (QCD) is another notable motivation for physics beyond the SM.
The smallness of the neutron electric dipole moment was explained by the Peccei-Quinn mechanism and the axion~\cite{Peccei:1977hh,Peccei:1977ur,Weinberg:1977ma,Wilczek:1977pj}, as the result of an approximately conserved global symmetry in QCD.
Due to the breaking of this symmetry, a light pseudo-Goldstone boson, the axion, arises and can be searched for via its non-renormalizable couplings to the SM gauge bosons and fermion content.
While the original solution to the strong CP problem is mostly excluded for axion masses of the order of MeV, it is still possible that light pseudoscalar particles associated with other global symmetries exist in this mass range.
Particles of this type that do not necessarily solve the CP problem and have their mass generated by means other than the QCD vacuum are known as axion-like particles (ALPs).

In this section, we study several different types of ALPs. The general form of the interactions we are interested in are,
\begin{equation}\label{eq:ALPgeneral}
   \mathscr{L}_{\rm ALP} \supset  \frac{1}{2}(\partial_\mu a) (\partial^\mu a) -  \frac{m_a^2}{2} a^2 + \frac{\partial_\mu a}{2 f_a} j^\mu,
\end{equation}
where the derivative coupling indicates this is indeed a pseudo-goldstone boson, and the current $j^\mu$ will take different forms in the sections that follow.

\begin{figure}[t]
    \centering
    \includegraphics[width=\columnwidth]{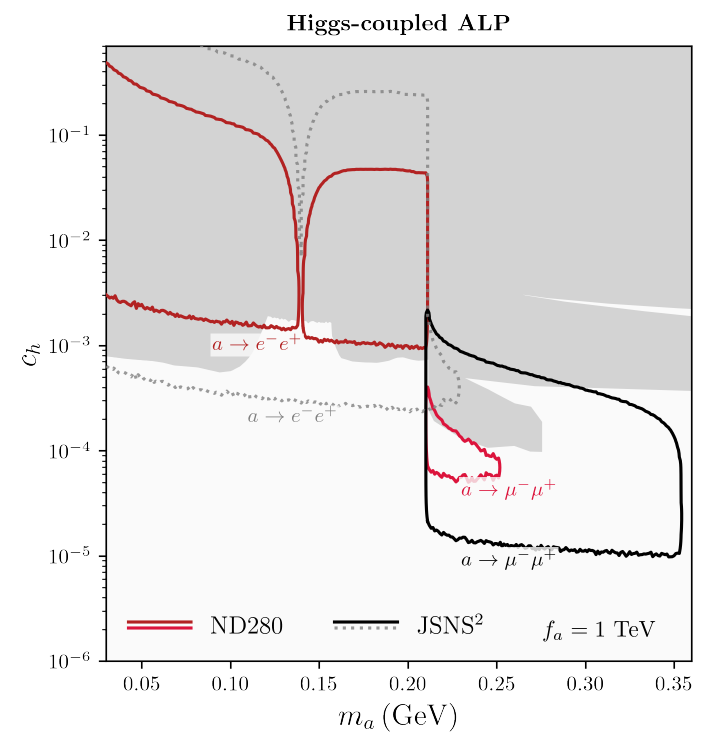}
    \caption{The projected sensitivities to ALPs produced in kaon decays are shown. Each color represents a specific decay final state, including $a_h\to e^+e^-$ and $a_h\to \mu^+\mu^-$. Solid lines indicate the sensitivities expected for the cleanest channels and dotted lines denote single-hit signals in \JSNSsquared. The current experimental bounds, extracted from Ref.~\cite{Coloma:2022hlv}, are shown in grey.
    % \mh{This should go in the main text: For all signals, we are setting the efficiency to 0.1.}\su{I have put it at the beginning of the Results section, for now I am assuming that for CCM and COHERENT we also use this $10\%$ }
    \label{fig:alp_sensitivity}
    }
\end{figure}

\begin{figure*}[t]
    \centering
    \includegraphics[width=\columnwidth]{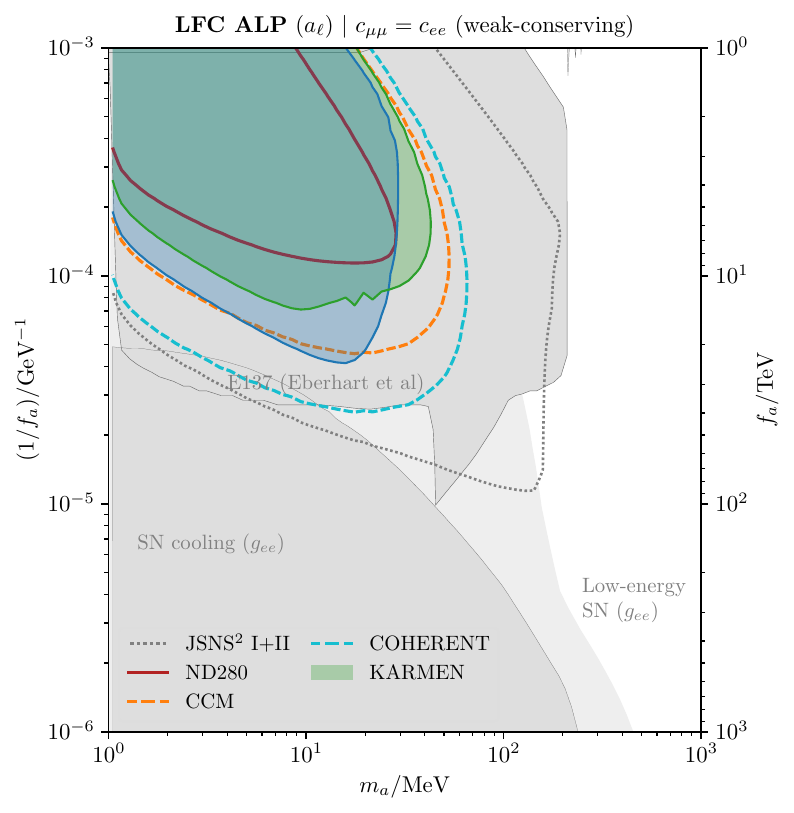}
    \includegraphics[width=\columnwidth]{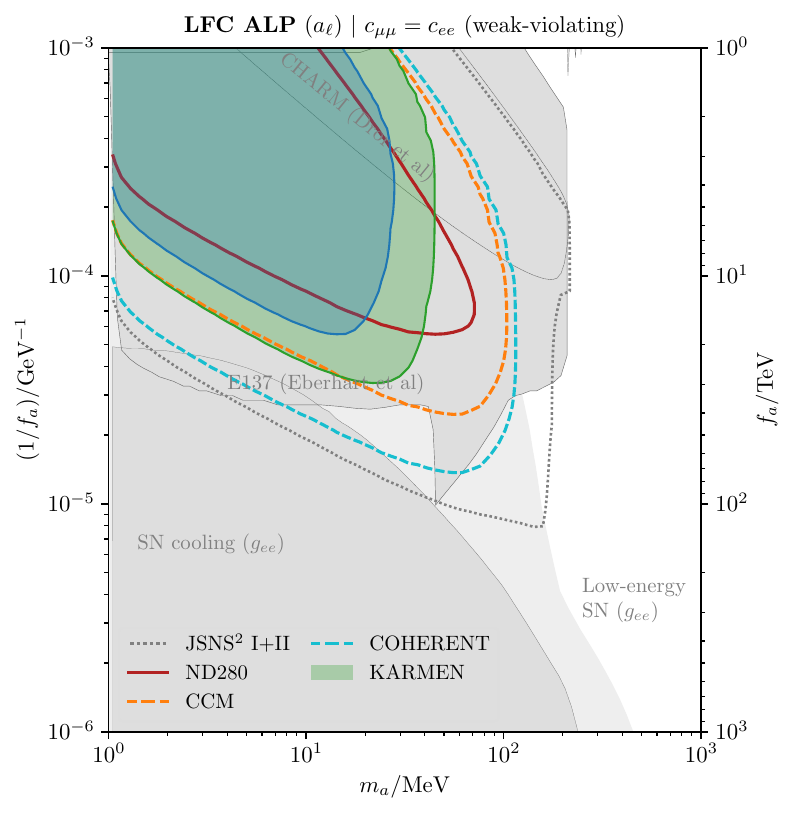}
    \caption{Projected sensitivities to leptophilic ALPs that couple universally to all leptons in a weak-conserving (left) and weak-violating (right) way. Solid lines indicate the sensitivities expected for the cleanest channels, dashed lines correspond to channels likely to suffer from significant background contamination, and dotted lines denote single-hit signals in \JSNSsquared. Also shown are constraints from KARMEN based on $\pi^+ \to e^+ \nu a_\ell$ production.     
   Other constraints include supernova limits~\cite{Fiorillo:2025sln} and other beam dumps~\cite{Eberhart:2025lyu,Altmannshofer:2022ckw}.
    \label{fig:wvalp_sensitivity}
    }
\end{figure*}

\begin{figure*}[t]
    \centering
    \includegraphics[width=\columnwidth]{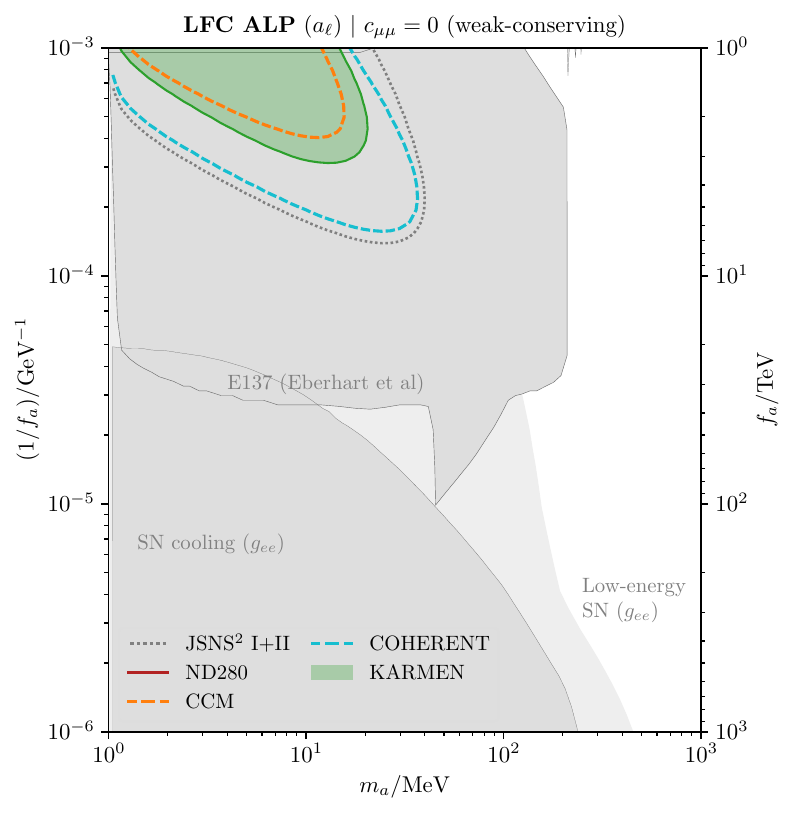}
    \includegraphics[width=\columnwidth]{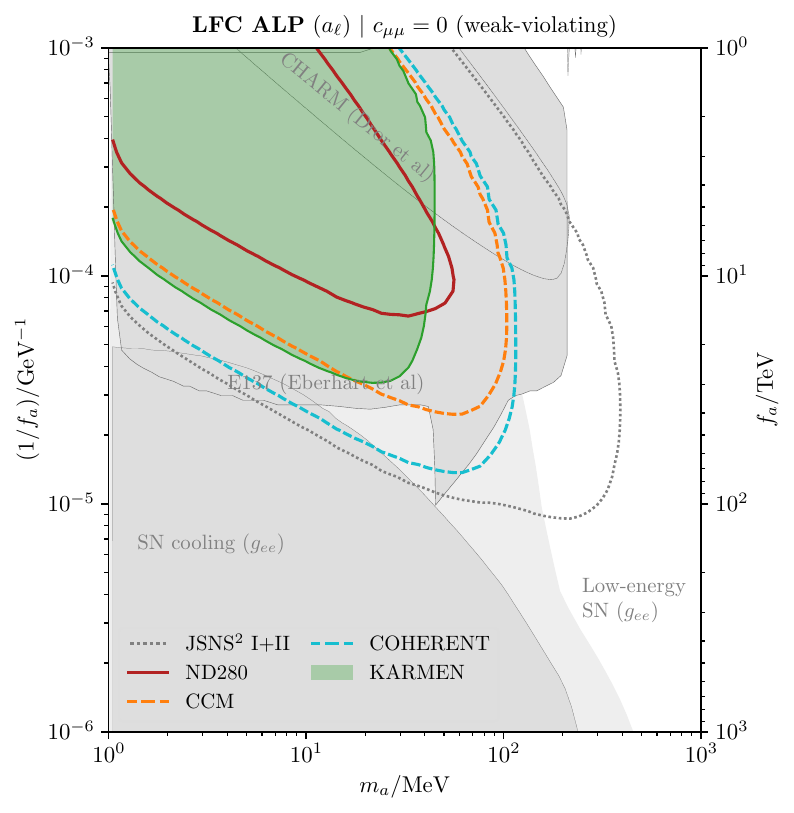}
    \caption{Projected sensitivities to leptophilic ALPs that couple only to electrons in a weak-conserving (left) and weak-violating (right) way. Solid lines indicate the sensitivities expected for the cleanest channels, dashed lines correspond to channels likely to suffer from significant background contamination, and dotted lines denote single-hit signals in \JSNSsquared. Also shown are constraints from KARMEN based on $\pi^+ \to e^+ \nu a_\ell$ production.     
   Other constraints include supernova limits~\cite{Fiorillo:2025sln} and other beam dumps~\cite{Eberhart:2025lyu,Altmannshofer:2022ckw}.
\label{fig:wvalp_sensitivity_electron}
    }
\end{figure*}

%%%%%%%%%%%%%%%%%%%%%%%%%%%%%%%%%%%
%\subsubsection{Gluon dominance}
%\label{sec:gluon_dominance}

%\begin{equation}\label{eq:ALPgluondominance}
%    \mathscr{L}_{\rm ALP-GG} \supset c_{GG}  \frac{a_G}{f_a} \frac{\alpha_S}{4 \pi} %G_{\mu\nu} \tilde{G}^{\mu\nu}
%\end{equation}

\subsection{Higgs-coupled ALP ($a_h$)}
\label{sec:higgscoupled_alp}

We start with the case where the ALP couples predominantly to SM fermions via the Higgs boson.
We refer to this ALP as $a_h$ and it couples to the current
\begin{equation}\label{eq:ALPlagrangian}
    j_{ h}^\mu =  2 \times c_h \left(H^\dagger i\overset{\leftrightarrow}{D^\mu} H \right) \longrightarrow  j_{h}^\mu = c_{ff} \sum_f \bar{f} \gamma^\mu\gamma_5 f,
\end{equation}
where $f_a$ is the ALP decay constant, $H$ is the SM Higgs doublet, and $f$ represents either charged leptons or quarks, with $c_{ff}=c_{h}$ for $f=d,s,b,e,\mu,\tau$ and $c_{ff} = -c_h$ for $f=u,c,t$.
The right-hand side of the arrow is obtained with a hypercharge rotation~\cite{Georgi:1986df}.
In general, the ALP can also couple to the SM via other gauge bosons. 
The coexistence of these different operators is indeed expected in UV completions of \cref{eq:ALPlagrangian} and, depending on their relative strengths, can lead to destructive interference to observables in low energies~\cite{Gavela:2019wzg}.
With this caveat, we proceed by considering only the coupling $c_h$.

The production of $a_h$ in pion and muon decays is suppressed by the small lepton masses.
However, similarly to the HPS, this ALP can be effectively constrained by FCNC kaon decays thanks to its sizable coupling to heavy quarks.
The production rate is given by
\begin{equation}
    \mathcal{B}(K^+ \to \pi^+ a_h) = \frac{|c_h|^2}{f_a^2}\frac{|k_{sd}|^2m_K^3}{64\pi} F_{\rm PS}\tau_{K^+},
\end{equation}
where the FCNC coefficient is approximately given by
\begin{equation}
    k_{sd} \simeq V_{td}^* V_{ts} \frac{g_2^2}{16 \pi^2} \frac{m_t^2}{v^2} \log\frac{\mu^2}{\Lambda^2},
\end{equation}
The phase space factor is $F_{\rm PS} \equiv \lambda^{1/2}(1,r_a^2, r_\pi^2)(1-r_\pi^2)^2$  with $r_a = m_a/m_K$ and $r_\pi = m_\pi/m_K$.
Here, $\mu$ is the renormalization scale, which we set to $2$GeV following\cite{Bauer:2021wjo}, and we take $\Lambda = f_a$ as the new-physics scale associated with the UV completion of \cref{eq:ALPlagrangian}.
Contrary to the HPS case, the FCNC rates have some UV dependence due to the non-renormalizable nature of the ALP couplings.
Numerically, $k_{sd} = V_{td}^* V_{ts} \times 8.2\times 10^{-3}$ for $f_a = 1$~TeV~\cite{Bauer:2017ris,Bauer:2020jbp,Gavela:2019wzg}.
% \mh{Agree? We were missing the CKM elements before I think}.\su{Agreed, yes it is with the CKM elements, there was a typo before}
% \mh{There is a subtlety about $\Lambda$. 
% Typically one thinks about the new physics scale as being $\Lambda= 4\pi f_a$ rather than $\Lambda=f_a$. For example, in ChPT, the breakdown of the EFT happens at $\Lambda= 4\pi f_\pi \sim 1$~GeV and not $\Lambda = f_\pi \sim 100$~MeV.}

In this minimal model, the ALP lifetime is also dictated by $c_h$.
The decay rate into leptons is given by
\begin{equation}\label{eq:alp_ll_decayrate}
    \Gamma_{a_h \to \ell^+ \ell^-} = |c_h|^2 \frac{m_a m_\ell^2}{8 \pi f_a^2}\sqrt{1 - 4r_\ell^2},
\end{equation}
where $r_\ell = m_\ell/m_a$.
At the one-loop level, $a_h$ can also decay to two photons. Since this is a subdominant branching ratio in the parameter space of interest, we neglect it here.

%%%%%%%%%%%%%%%%%%%%%%%%%%%%%%%%%%%%%%%%%%%%%%%%%%%%%%%%%%%%%%%%%%%%%%
\subsection{Leptophilic Axion-Like-Particles ($a_{\ell}$)}
\label{sec:wvalps}
%%%%%%%%%%%%%%%%%%%%%%%%%%%%%%%%%%%%%%%%%%%%%%%%%%%%%%%%%%%%%%%%%%%%%%

We now consider ALPs coupled exclusively to leptonic currents.
The most general parameterization of the ALP couplings to leptons reads
\begin{align}\label{eq:alp_leptonic_current}
   j^\mu_{\ell} &=  \sum_{i,j}^{e,\mu,\tau} c_{ij}^L \overline{\ell^i_L} \gamma^\mu \ell^j_L + c^\nu_{ij} \overline{\nu_L^i} \gamma^\mu \nu_L^j + c_{ij}^R \overline{\ell_R^i} \gamma^\mu \ell_R^j.
\end{align}
The choice of the coupling matrices $c^L$, $c^\nu$, and $c^R$ will dictate the dominant production and decay modes of the ALPs.
In what follows, we discuss a few different regimes of interest and narrow down \cref{eq:alp_leptonic_current} into popular benchmarks.

The coupling matrices $c^{L,R,\nu}$ are Hermitian in flavor space, which for simplicity we assume to be real, so $c^{L,R,\nu}_{ij} = c^{L,R,\nu}_{ji}$.
Note that in all generality, $c^L$ couplings are not necessarily related to $c^\nu$ couplings, as one might naively impose to respect SU$(2)_L$.
To see this, one may work instead with the dimension-7 operators $(\partial_\mu a) \left({\overline{L_i H} \gamma^\mu H L_j}\right)$ and $(\partial_\mu a) \left({\overline{L_i H^\dagger} \gamma^\mu H^\dagger L_j}\right)$.
Separating couplings in this fashion allows us to investigate the so-called weak-violating ALP models of~\cite{Altmannshofer:2022ckw}, where the helicity suppression in meson decays to ALPs of the type $\pi^+\to e^+ \nu_e a_\ell$ is lifted.
When $c^L_{ij} = -c^\nu_{ij}$, the underlying operators are simply $(\partial_\mu a) \left(\overline{L_i}\gamma^\mu L_j\right)$ and we refer to the model as a weak-preserving one.

In the most general case we have both lepton-flavor-violating (LFV) couplings ($i\neq j)$ and lepton-flavor-conserving (LFC) couplings ($i = j$), with a total of $3\times 6 = 18$ constants to choose from.
In particular, if $c_{e\mu}^{L,R} \neq 0$, the production of the ALP can be significantly enhanced due to 2-body muon decays such as $\mu^+ \to e^+ a_\ell$.
In LFC scenarios, $c_{ij}^{L,R,\nu} = \delta_{ij} \times c_j^{L,R,\nu}$, so a total of $9$ couplings may be non-zero.
Imposing a further constraint of lepton-flavor-universality (LFU), one would then have $c_{ij}^{L,R,\nu} = \delta_{ij} \times c^{L,R,\nu}$, reducing the number of couplings to only $3$.
Therefore, for a LFU and LFC ALP, one must only specify the relationship between three universal couplings: $c^L, c^R$, and $c^\nu$.

For simplicity, we only specify the phenomenologically relevant couplings and set all remaining terms to zero.
As we will see, the different choices correspond to scenarios where $\pi^+$, $K^+$, and $\mu^+$ DAR can provide interesting sensitivity to the parameter space of leptophilic ALPs.

\subsubsection{Lepton-flavor conserving (LFC)}
Within LFC scenarios, we consider two representative cases:
\begin{itemize}
    \item Weak-conserving (WC): 
    \begin{equation}
        c_{ii}^L= -c_{ii}^R, 
        \qquad c^\nu_{ii} = -c^L_{ii}.
    \end{equation}

    \item Weak-violating (WV):
    \begin{equation}
        c_{ii}^L= -c_{ii}^R, , 
        \qquad c^\nu_{ii} = 0.
    \end{equation}
    Because these leptonic couplings are not explicitly SU$(2)_L$-invariant,\footnote{Gauge invariance can be recovered with appropriate UV completions, such as by coupling the ALP to right-handed neutrinos or vector-like leptons, which in turn induce such ``weak-violating'' couplings through mixing in the lepton sector~\cite{Altmannshofer:2022ckw}.} the ALP production can lift the helicity suppression in meson decays such as $\pi^+ \to e^+ \nu_e a$~\cite{Altmannshofer:2022ckw}, constituting an important production mode of this kind of ALP at spallation sources.
    While these models pertain to more specific UV completions, they represent a lamppost for experiments dealing with copious amounts of charged mesons.
\end{itemize}

The relevant production modes for leptophilic ALPs in the LFC scenarios are summarized in \cref{app:branchings}.

%%%%%%%%%%%%%%%%%%%%%%%%
%\subsubsection{LFC weak-conserving}

%This case corresponds to the choice:
%\begin{equation}
%    c_{ee}^L= c_{\mu \mu}^L = -c_{ee}^R= -c_{\mu \mu}^L \text{ and } c^\nu_{ii} = -c^L_{ii}.
%\end{equation}

%%%%%%%%%%%%%%%%%%%%%%%%
%\subsubsection{LFC weak-violating}
%This case corresponds to the choice:
%\begin{equation}
%    c_{ee}^L= c_{\mu \mu}^L = -c_{ee}^R= -c_{\mu \mu}^L \text{ and }c^\nu_{ii} = 0
%\end{equation}
%Because these leptonic couplings are not explicitly SU$(2)_L$-invariant~\footnote{Gauge invariance can be recovered with appropriate UV completions, such as by coupling the ALP to right-handed neutrinos or vector-like leptons, which in turn induce such ``weak-violating" couplings through mixing in the lepton sector~\cite{Altmannshofer:2022ckw}.}, the ALP production can lift the helicity suppression in meson decays such as $\pi^+ \to e^+ \nu_e a$~\cite{Altmannshofer:2022ckw}, constituting an important production mode of this kind of ALP at spallation sources.
%While these models pertain to more specific UV completions, they represent a lamppost for experiments dealing with copious amounts of charged-mesons.

%%%%%%%%%%%%%%%%%%%%%%%%
\subsubsection{Lepton-flavor violating (LFV)}

This case corresponds to choosing non-vanishing off-diagonal components in the coupling matrix. For simplicity, we set all tau couplings to zero and make a weak-conserving choice for the remaining couplings,
\begin{equation}
c_{ij}^L = -c^R_{ij} \text{ and } c^\nu_{ij} = -c^L_{ij}. 
\end{equation}
Flavor universality is not necessarily required and, as we will see, the non-universal case with a great hierarchy between off- and on-diagonal couplings is the most advantageous for LLP-DIF signals.

If $c_{e\mu}$ LFV couplings coexist with $c_{ee}$ LFC couplings, then for light ALPs with $m_a < m_\mu - m_e$, the production $\mu^+ \to e^+ a_\ell$ followed by a displaced $a_\ell \to e^+e^-$ or $a_\ell \to \gamma \gamma$ decay provides a compelling discovery channel.
These signatures can be advantageous in comparison to searches for exotic muon decays, such as $\mu \to e a_\ell^{\rm inv}$, thanks to the sheer number of muon decays at spallation sources.
In addition, if $c_{e\mu}/c_{ee} < 1$, then the decay-in-flight signatures can become advantageous, as the limits on the decay constant scale as $f_a^{-1} \propto \sqrt{c_{e\mu} c_{ee}}$ rather than $f_a^{-1} \propto c_{e\mu}$.
Similarly, direct observation of rare decays like $\mu \to e \gamma$ and $\mu^+ \to e^+e^+e^-$ are also not competitive despite the similar scaling with the new physics couplings thanks to the smaller number of muons involved.

The most relevant decay in LFV scenarios is two-body muon decays with a total branching ratio of
\begin{equation}
    \Gamma(\mu^+ \to e^+ a_{\ell}) \simeq (|c_{e \mu}^L|^2 + |c_{e \mu}^R|^2) \frac{m_\mu^3}{32 \pi f_a^2} f(r_e, r_a),
\end{equation}
where $f(r_e, r_a) = (1 + r_e)^2\left[(1 -r_e)^2 - r_a^2\right]\lambda^{1/2}(1,r_e, r_a)$ with $r_i = m_i^2/m_\mu^2$.
These two-body decays dominate the total ALP emission, providing an approximately mono-energetic ALP flux.
The decay of the ALP is then, similarly to before, just given by
\begin{align}
\Gamma(a_\ell \to e^+e^-) = \left(|c_{ee}^L|^2 + |c_{ee}^R|^2\right)\frac{m_a m_e^2}{16\pi f_a^2}\left(1-\frac{4 m_e^2}{m_a^2}\right)^{1/2}.
\end{align}
In LFV scenarios, the ALP could in principle decay to final states such as $a_\ell \to \mu^\pm e^\mp$, but because the only relevant production channel at spallation sources in those cases will be $\mu^+ \to e^+ a_\ell$, the only relevant decay channel will be $a \to e^+ e^-$.
We do not consider the case where $a_\ell \to \gamma \gamma$ can dominate the ALP branching ratio as it requires suppressed direct coupling to electrons which, in the LFV model, will already be excluded by missing-energy searches in $\mu^+ \to e^+ a_\ell^{\rm inv}$.
% In addition to that, ALPs can also be emitted via the three-body process $\mu^+ \to e^+(p_1) + \gamma(p_2) + a_{\rm fv}(p_3)$,
% \begin{equation}
%     \frac{\mathcal{B}(\mu^+ \to e^+ \gamma a_{\rm fv})}{\dd x_{12} \dd x_{23}} = c_{e\mu}^2\frac{\alpha}{128\pi^2}\frac{m_\mu^3}{x_{12}} \frac{g(x_{12}, x_{23})}{(x_{a_{\rm fv}}-x_{12}-x_{23})^2},
% \end{equation}
% where $x_{ij} = (p_i+p_j)^2/m_\mu^2$, $x_{a_{\rm fv}} = m_a^2/m_\mu^2$ and $g(x,y) = x_{a_{\rm fv}}^3 - y^2(x+y) -x_{a_{\rm fv}}^2(2+x+y) + 2(x+y)(2x+y) - 2(4x+y) + x_{a_{\rm fv}}(2+4x+y^2)$.

\begin{figure*}[t]
    \centering
    \includegraphics[width=\columnwidth]{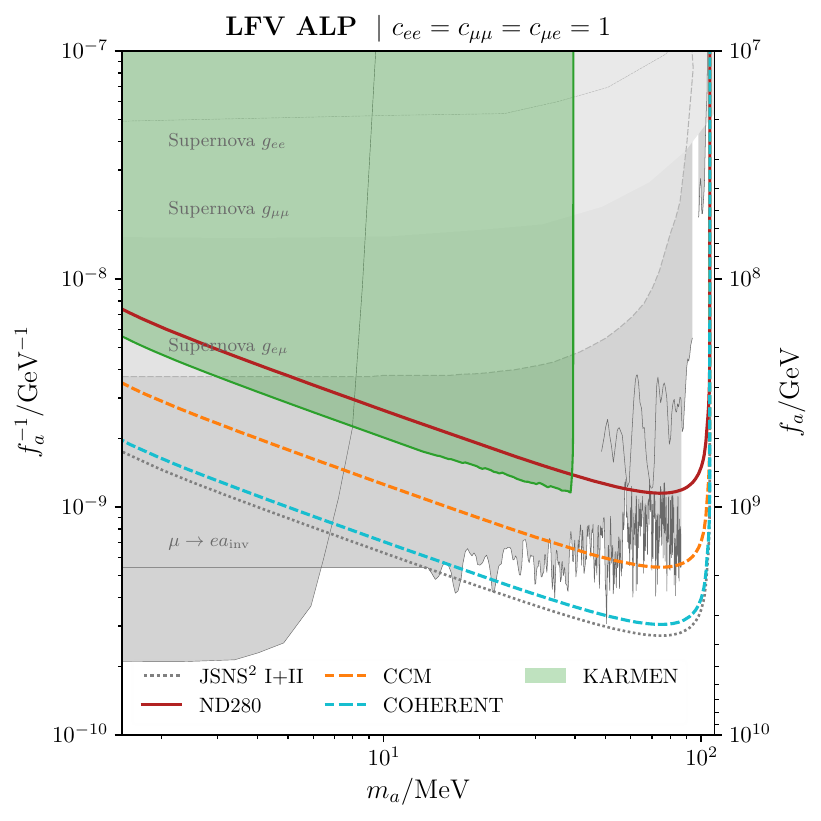}
    \includegraphics[width=\columnwidth]{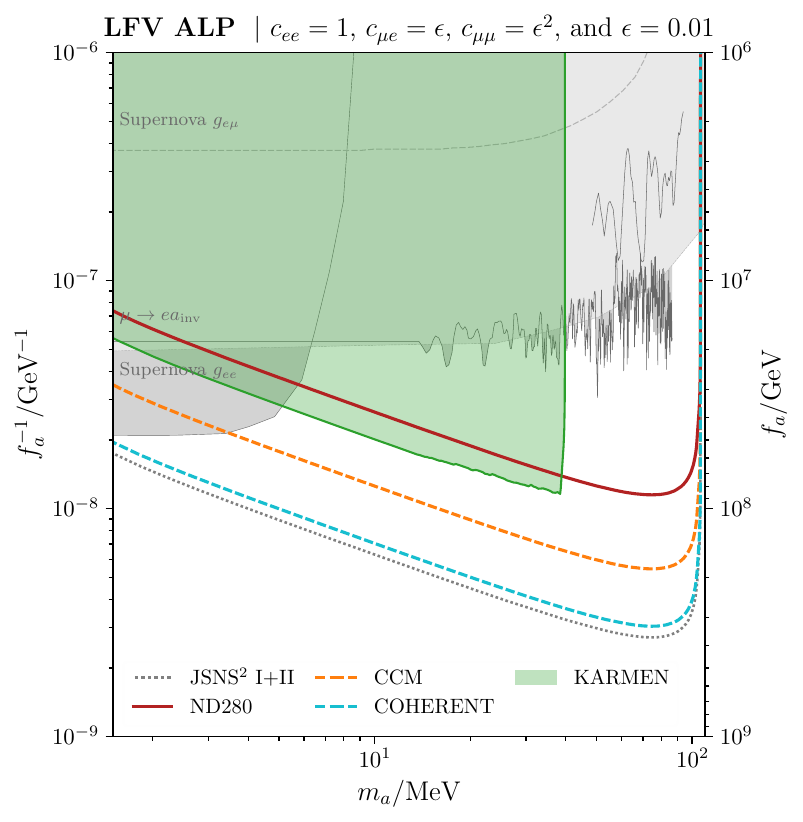}
    \caption{
    Projected sensitivities to lepton flavor violating ALPs. Solid lines indicate the sensitivities expected for the cleanest channels, dashed lines correspond to channels likely to suffer from significant background contamination, and dotted lines denote single-hit signals in \JSNSsquared. Also shown are the bounds derived from KARMEN data.
    On the left, we show the case with $c_{ee} = c_{e\mu} = c_{\mu \mu} = 1$ and on the right we fix $c_{ee} = 1$, $c_{\mu e}=10^{-2}$, $c_{\mu \mu}=10^{-4}$.
    Constraints from $\mu \to e a$ and from supernovae are shown as light gray shaded areas.
    Because we only include $\mu^+ \to e^+ a$ decays at production, no bound from LSND is derived as $E_a^{\rm min} > 50$~MeV, the highest energy bin of LSND data. 
    \label{fig:fvalp_sensitivity}
    }
\end{figure*}

\subsection{Sensitivities}

The results for the Higgs-coupled ALPs are shown in \cref{fig:fvalp_sensitivity,}.
In this scenario and for the considered mass range, the dominant production channel is the kaon decay $K \to \pi a$. Once produced, the ALP can decay into three visible final states: $e^{+}e^{-}$, $\mu^{+}\mu^{-}$, and $\gamma\gamma$. Hadronic modes are not permitted in this range---specifically, the decays $a \to \pi\pi$ and $a \to \pi^0\gamma$ are forbidden by CP and C conservation, respectively, while $a \to 3\pi$ is kinematically inaccessible. As in the HPS case, only the JSNS facility provides a sufficiently energetic kaon flux among the currently operating spallation sources, so we focus on the sensitivity of \JSNSsquared. The results are presented separately for each visible channel. At low masses, decay-in-flight searches at ND280 are competitive with existing kaon decay limits that do not rely on reconstructing the ALP decay. In our case, the sensitivity scales with the square of the mixing to the Higgs, and the observable signal with the fourth power, making improvements challenging in this region. Only if \JSNSsquared were able to reduce the background for the single-hit signal ($e^+e^-$) to a negligible level would improvement be possible. For $m_a > 2m_\mu$, the $\mu^+\mu^-$ mode provides a particularly clean signature at \JSNSsquared: the resulting triple-hit signal offers strong background rejection, potentially surpassing current bounds by more than an order of magnitude.

The leptophilic LFC ALP results for both weak-conserving (left panel) and weak-violating (right panel) scenarios are shown in \cref{fig:wvalp_sensitivity} for the flavor-universal case. In the weak-conserving scenario, due to the helicity suppression of $M^{+} \rightarrow e^{+} \nu_e a$ ($M = \pi, K$), the production is dominated by $M^{+} \rightarrow \mu^{+} \nu_{\mu} a$ and $\mu^{+} \rightarrow e^{+} \nu \nu a$. In contrast, in the weak-violating scenario (right panel) the helicity suppression is lifted, and $\pi^{+} \rightarrow e^{+} \nu_e a$ becomes an important production channel. In both cases, although \JSNSsquared and COHERENT could potentially improve upon laboratory bounds, the very strong low-energy constraints from supernova observations remain more stringent. Note that the sensitivity of \JSNSsquared drops abruptly around 200~MeV, corresponding to the opening of the $a_\ell \rightarrow \mu^+ \mu^-$ channel, which renders the ALP too short-lived to be detected. Motivated by this, we also show results for the case where the ALP couples only to electrons.

\cref{fig:wvalp_sensitivity_electron} shows this electron-coupled ALP scenario.
In the WC scenario (left panel), the production is then driven by $M^{+} \rightarrow e^{+} \nu_e a$, which remains helicity suppressed and thus leads to a much weaker sensitivity. 
On the other hand, in the WV scenario (right panel), once this suppression is lifted and the only possible ALP decay is $a \to e^+ e^-$, the reach of the (very optimistic) single-hit sensitivity of \JSNSsquared extends beyond the low-energy supernova constraints. 
Note that the current constraint remains the same across all the different LFC scenarios considered, since both beam-dump and supernova bounds are driven by the electron coupling (the only exception are the CHARM limits derived in \cite{Altmannshofer:2022ckw}).

\cref{fig:fvalp_sensitivity} shows the LFV ALP results for both a flavor-universal (left panel) and an electron-dominated (right panel) scenario. 
In the flavor-universal case, the strong bounds from rare $\mu$ decays imply that the possible improvement from spallation sources is only mild. 
The advantage of rare $\mu$ decays is that, for small couplings, the sensitivity scales as $\frac{c_{\mu e}^2}{f_a^2}$, allowing strong constraints even with more limited statistics. 
In contrast, the decay-in-flight sensitivity scales as $\frac{c_{\mu e}^2 c_{ee}^2}{f_a^4}$. 
This makes $\mu$ decays in flight particularly advantageous whenever the electron coupling dominates. 
On the right panel, we consider a texture where $c_{\mu e}$ is suppressed by two orders of magnitude. 
In this case, the rare $\mu$ decay bounds weaken by two orders of magnitude, whereas the decay-in-flight sensitivity only worsens by one. 
For this scenario, spallation sources are therefore expected to improve substantially over existing bounds.

%%%%%%%%%%%%%%%%%%%%%%%%%%%%%%%%%%%%%%%%%%%%%%%
\section{Conclusions}
\label{sec:conclusions}

We surveyed modern neutrino detectors at spallation sources and studied their sensitivity to the decays in flight of long-lived particles (LLP).
The high-intensity proton beams produce unparalleled numbers of pion, kaon, and muon decay at rest which lead to a potential source of LLPs.
Considering current and future neutrino detectors, we identified several opportunities to cover new parameter space in models of heavy neutral leptons, dark scalars, and axion-like particles. 
The new physics signatures are characterized by high-energy charged particles or photons from the decay in flight of LLPs into dilepton, diphoton, or dipion final states.
Ultimately, our goal is to motivate experimental collaborations to carry out dedicated sensitivity studies to determine how close to the optimistic case scenario presented here they would be.

Thanks to the fortuitous location of several detectors at the J-PARC campus, JSNS can be used as a source of LLPs with varied detection technologies: the gaseous argon TPCs of ND280(+), the Cherenkov and scintillation detectors of JSNS$^2$, and the large volume ECAL systems of KOTO.
ND280+ stands out as a particularly clean environment where a low-background search could be performed if the detector can trigger on such events.

At Oak Ridge, the suite of COHERENT detectors also offers new opportunities thanks to the low neutron backgrounds in the neutrino alley.
We find that the combination of detectors like the D2O and H2O modules, NAIvETE, and COH-Ar-750 can amount to a total target mass of a few tons when combined.
Clearly more dedicated studies of the background rejection level should be carried out, but with optimistic background assumptions in 3 years of SNS operation, we find that COHERENT can probe new parameter space in several models.
New detectors have also been proposed for the first and second target stations at SNS~\cite{Asaadi:2022ojm}.
A new generation of ``discovery-level" ton-sized detectors inside or outside the SNS buildings could push the sensitivities shown here even further.

At the Los Alamos Neutron Science Center (LANSCE), the Coherent Captain-Mills (CCM) experiment also presents a compelling opportunity. The 5-ton liquid argon detector is well-positioned to search for LLPs with masses below the pion threshold, produced in pion and muon decays. Its sensitivity to final states such as $e^+e^-$ and $\gamma\gamma$ would be competitive with other facilities, provided that beam-related backgrounds can be effectively mitigated. The ongoing upgrade to the CCM200 detector, featuring improved liquid argon purity, enhanced shielding, and significantly more data, is a crucial step toward realizing this potential. The CCM collaboration is already pursuing searches for dark matter and axion-like particles, and our work demonstrates that a dedicated analysis for decay-in-flight signatures could broaden its physics reach.

We conclude by emphasizing that spallation neutron sources provide long-term, well-controlled environments in which to look for new particles.
As facilities in high demand for material science, they rely on accelerator-driven spallation rather than critical chain reactions for neutron production, being more reliable and less temperamental than reactor sources.
Therefore, the program we discuss here is likely to stay relevant for many years to come.
In the near to intermediate future, one can envision large detectors exposed to LANSCE, SNS, or JSNS, improving the sensitivities shown here even further.
Future projects include the SHiNESS detector at the ESS~\cite{Abele:2022iml,Soleti:2023hlr}, the water-based liquid scintillator detector EOS~\cite{Anderson:2022lbb} or the few-ton-scale PROSPECT-II detector~\cite{PROSPECT:2021jey} at SNS.

\acknowledgments{
We would like to thank Carlos Arg{\"u}elles and Pilar Coloma for their involvement in the early stages of this work.
We thank Jeff Dror and Wolfgang Altmannshofer for discussions on weak-violating axion-like-particles, Joshua Spitz and Teppei Katori for discussions on the capabilities of JSNS$^2$ and ND280, and Bryce Littlejohn for discussions on PROSPECT-II and EOS at SNS. 
The work of MH was partly supported by the Neutrino Theory Network Program Grant \#DE-AC02-07CHI11359 and the US DOE Award \#DE-SC0020250. 
We acknowledge the use of AI tools for the graphical design of the experimental set-ups.
}

\appendix

%%%%%%%%%%%%%%%%%%%%%%%%%%%%%%%%%%%%%%%%%%%%%%%%%%
\section{Limits from LSND}
\label{app:LSND}

\begin{figure}
    \centering
    \includegraphics[width=0.49\textwidth]{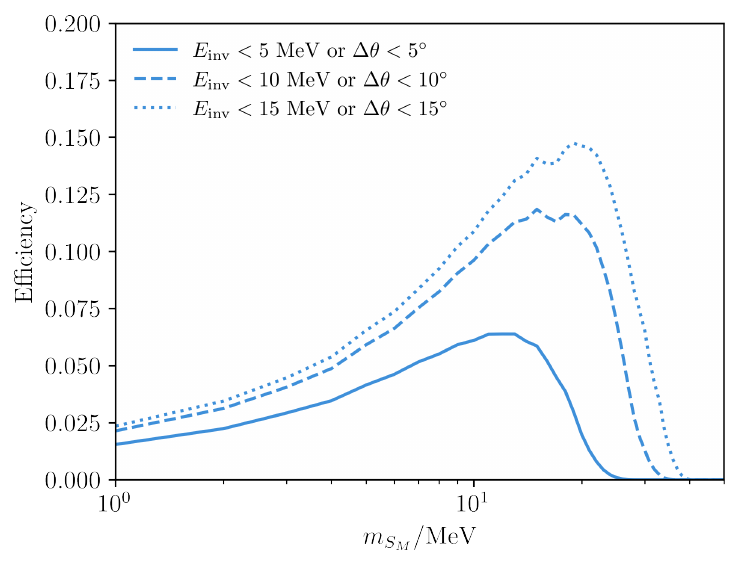}
    \caption{The resulting efficiencies for different selection criterion on $S_M \to e^+e^-$ or $S_M \to \gamma\gamma$ pairs to mimic a $\nu-e$ scattering event in LSND in the MPS model.
    All curves include the additional selection in \cref{eq:sig_selection_LSND}.}
    \label{fig:LSDN_efficiencies}
\end{figure}

When setting LSND limits and motivated by Ref.~\cite{Ema:2023buz} we assume that $e^+e^-$ or $\gamma\gamma$ pairs can no longer be distinguished as two separate particles under the following three scenarios:
\begin{subequations}
\begin{align}
    E_{\rm inv} &< 5 \text{ MeV or } \Delta \theta < 5^\circ,
    \\
    E_{\rm inv} &< 10 \text{ MeV or } \Delta \theta < 10^\circ,
    \\
    E_{\rm inv} &< 15 \text{ MeV or } \Delta \theta < 15^\circ.
\end{align}
\end{subequations}
which yield the efficiencies shown in~\cref{fig:LSDN_efficiencies} for the MPS model. 
In the main text, assume the first case as this is the most conservative choice.
We ensure that the resulting object obeys 
\begin{equation}\label{eq:sig_selection_LSND}
    \cos \theta > 0.9 \text{ and } 18 \text{ MeV } < E_{\rm vis} < 50 \text{ MeV}.
\end{equation}
after smearing the individual energies of the particles by a Gaussian resolution of $\sigma_E/E = 0.7\%$. We further apply an additional $50\%$ efficiency factor to account for extra fiducialization cuts and for the rejection of contamination from electrons produced in cosmic-ray interactions, following Table 1 of Ref.~\cite{LSND:2001akn}.

%%%%%%%%%%%%%%%%%%%%%%%%%%%%%%%%%%%%%%%%%%%%%%%%%%
\section{Limits from KARMEN}
\label{app:KARMEN}

\begin{figure*}[t]
    \centering
    \includegraphics[width=0.49\textwidth]{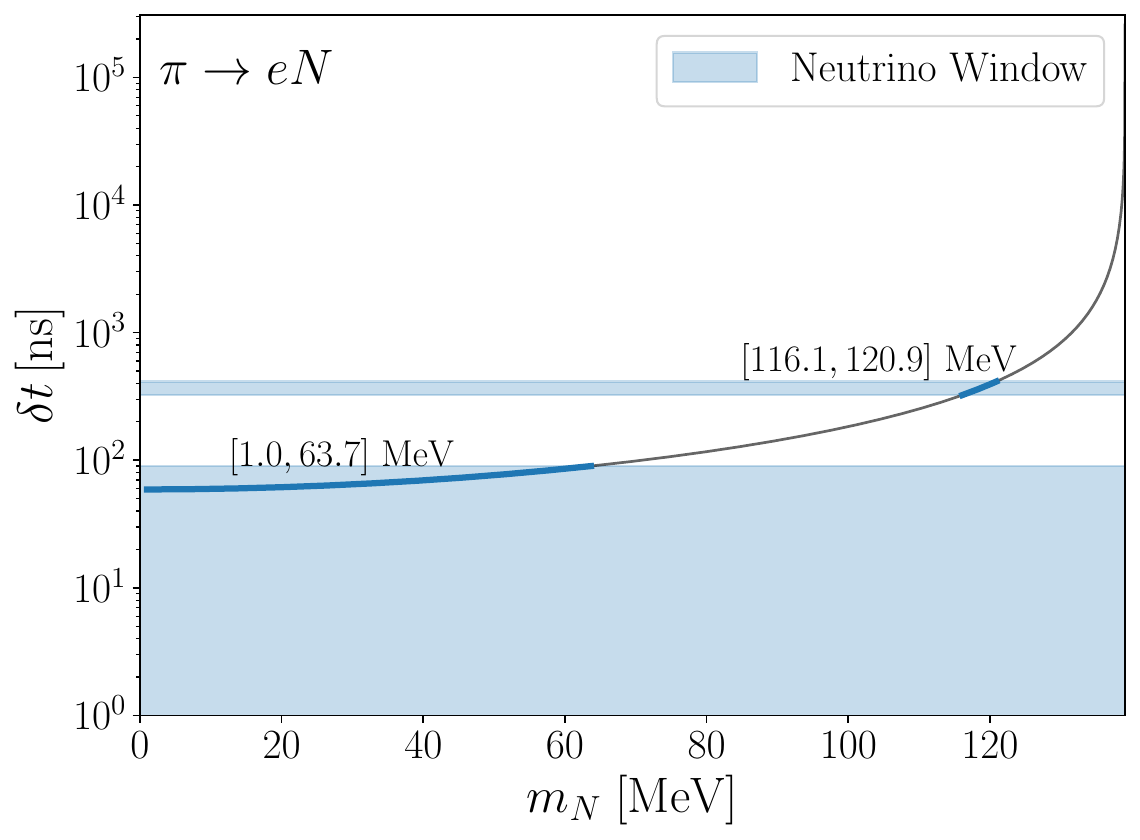}
    \includegraphics[width=0.49\textwidth]{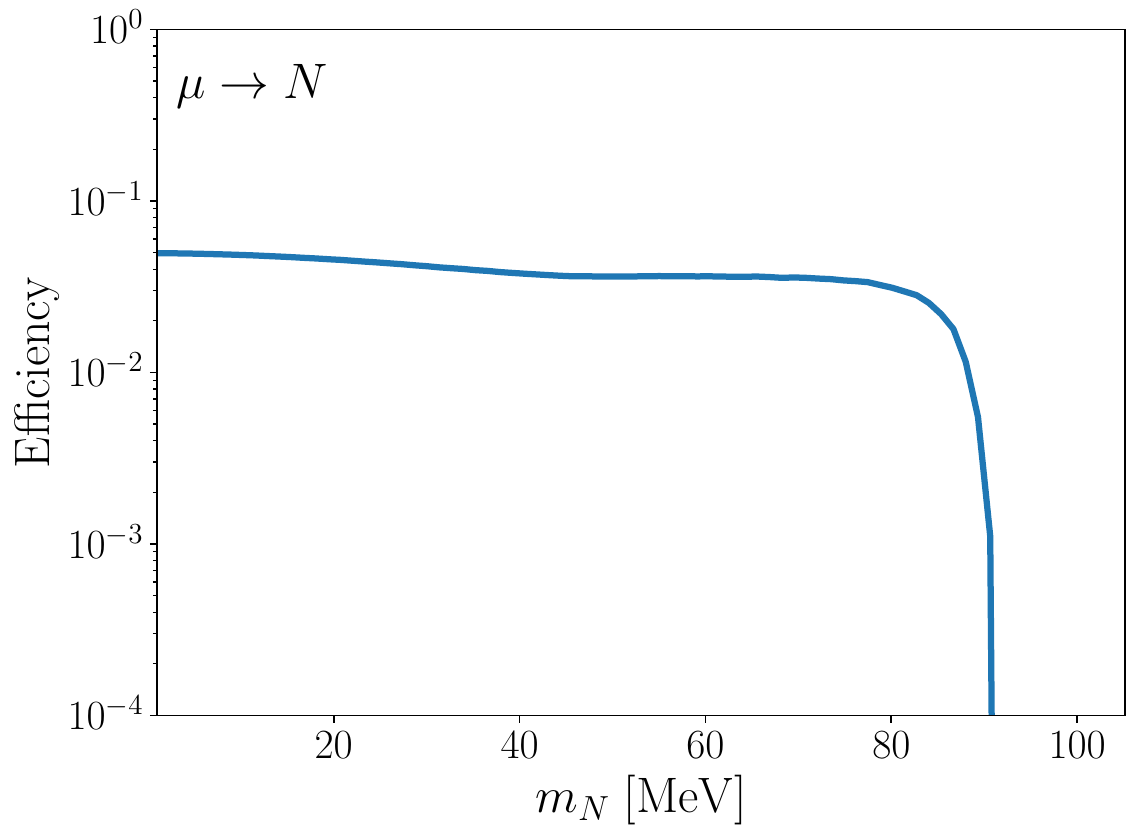}
    \caption{
    Left) Time delay $\delta t$ for a HNL produced in $\pi \to e N$ as a function of $m_N$.  
    The shaded bands correspond to the neutrino time window at KARMEN, \cref{eq:neutrino_window}.
    Right) Fraction of HNL events from $\mu$ decay that fall within the KARMEN neutrino time window as a function of $m_N$.
    \label{fig:time_delay}}
\end{figure*}

To set the KARMEN limits, we use the results of Ref.~\cite{KARMEN:1998xmo}, 
where the NC cross section of  
\begin{equation}
    \nu_\mu + {}^{12}\mathrm{C} \ \rightarrow\ \nu_\mu + {}^{12}\mathrm{C}^\ast(15.1\ \mathrm{MeV}),
\end{equation}
was measured for neutrinos produced in $\pi^+$DAR. 
The ISIS source employs a double-pulse proton beam, with each pulse having a width of 
$100\ \mathrm{ns}$ and separated by $324\ \mathrm{ns}$, at a repetition rate of $50\ \mathrm{Hz}$. 
For each double pulse, the ``neutrino window'' is defined as 
\begin{equation}\label{eq:neutrino_window}
0 < t - t_0 < 90\ \mathrm{ns} \text{ and }    324\,\mathrm{ns}< t - t_0 < 414\,\mathrm{ns},
\end{equation}
where $t_0 =0$ corresponds to the start of each bunch; outside these intervals, events 
are dominated by beam-related neutron backgrounds. 

The published KARMEN data cover visible energies from $10$--$100\ \mathrm{MeV}$.  
Above $15\ \mathrm{MeV}$, the observed number of events is very close to zero.  
In \cref{fig:Sm_with_data}, we show the number of events as a function of visible energy in the neutrino window. 
To set bounds on LLPs decaying into $e^+e^-$ and $\gamma\gamma$ pairs, we use the Feldman--Cousins limit~\cite{Feldman:1997qc} for an expected background of five events, which at $90\%$ C.L. corresponds to a cut of $n_{\text{events}} = 4.99$ above $15\ \mathrm{MeV}$ in visible energy. 

% \begin{figure}
%     \centering
%     \includegraphics[width=0.49\textwidth]{Figs/Karmen_data.pdf}
%     \caption{Energy distribution of the events in the KARMEN neutrino window (see text), after background subtraction, extracted from Fig.~4 of Ref.~\cite{KARMEN:1998xmo}.  
%     \mg{What does it mean to assign an uncertainty of five events? --- For the remaining energy bins (up to 100~MeV) not shown, we conservatively assume zero observed events and assign an uncertainty of five events.}}
%     \label{fig:Karmen_data}
% \end{figure}

%For the analysis, we use an energy-binned $\chi^2$:
%\begin{equation}
%    \chi^2 = \sum_{i=1}^{N} \fra{\left( n_{\mathrm{X}, i} - n_{\mathrm{data}, i} \right)^2}{\sigma_i^2} \,,
%\end{equation}
%where the bins span $15$--$100\ \mathrm{MeV}$ in visible energy, $n_{\mathrm{X}, i}$ denotes the predicted number of $X_{\text{LLP}}$ events in the $i$-th bin occurring within the neutrino window, and $\sigma_i$ is the corresponding uncertainty.

When $X_{\text{LLP}}$ is produced from pion decay, it acquires an additional time delay  according to \cref{eq:time_delay}.
As an illustration, the left panel of \cref{fig:time_delay} shows $\delta t$ for a HNL produced in $\pi^+ \to e^+ N$ decays as a function of its mass $m_N$.
It also shows the region of masses where the HNL arrival time lies inside the neutrino time window.  

For production via $\mu$ decay, the situation is different: the muon lifetime is $\tau_\mu \simeq 2.2\ \mu\mathrm{s}$, much longer than the bunch duration, so most decays occur well outside the neutrino windows.  
As a result, the fraction of signal events from muon decays that fall inside the neutrino window is suppressed.  
This efficiency, as a function of $m_N$, is shown on the right panel of \cref{fig:time_delay}.

We have estimated the time cut efficiency for standard neutrinos from pion decay to be approximately $40\%$, in order to reproduce the combined time and energy cut efficiency of $16\%$ reported in Ref.~\cite{KARMEN:1998xmo}. We therefore apply an additional overall efficiency factor of $40\%$ in both pion- and muon-induced cases.

%%%%%%%%%%%%%%%%%%%%%%%%%%%%%%%%%%%%%%%%%%%%%%%%%%
\section{Branching ratios}
\label{app:branchings}

\begin{figure*}[t]
    \centering
    \includegraphics[width=\columnwidth]{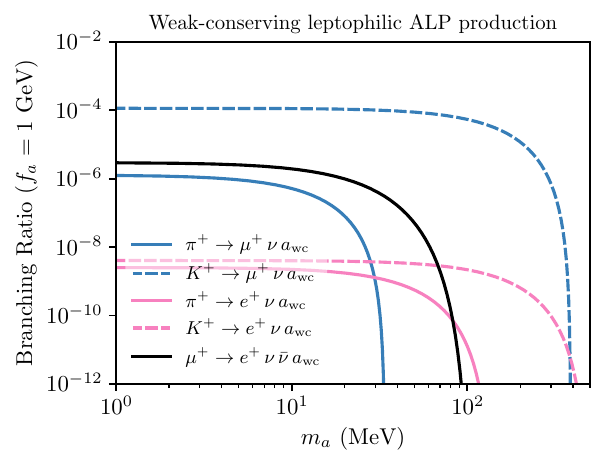}
    \includegraphics[width=\columnwidth]{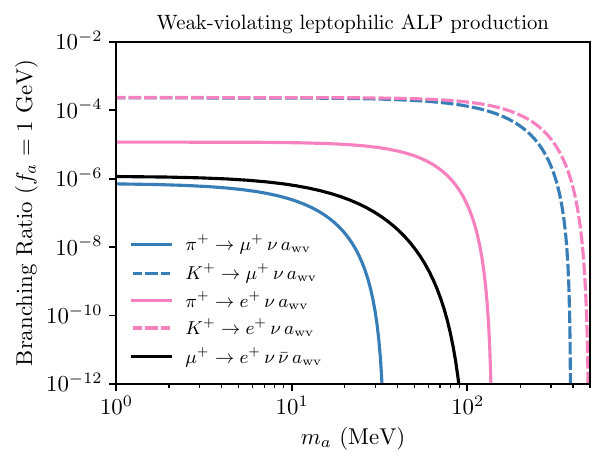}
    \caption{The pion, kaon, and muon branching ratios into leptophilic ALPs $a_\ell$ as a function of the ALP mass in LFC models of the weak-conserving (left) and weak-violating (right) type.
    \label{fig:branchings_lalp}
    }
\end{figure*}

\begin{figure*}[t]
    \centering
    \includegraphics[width=\columnwidth]{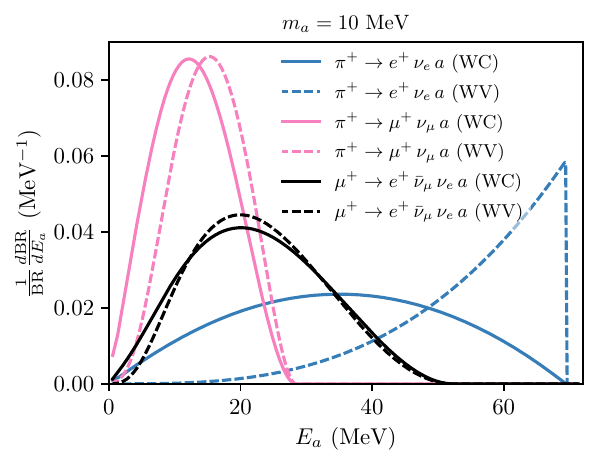}
    \includegraphics[width=\columnwidth]{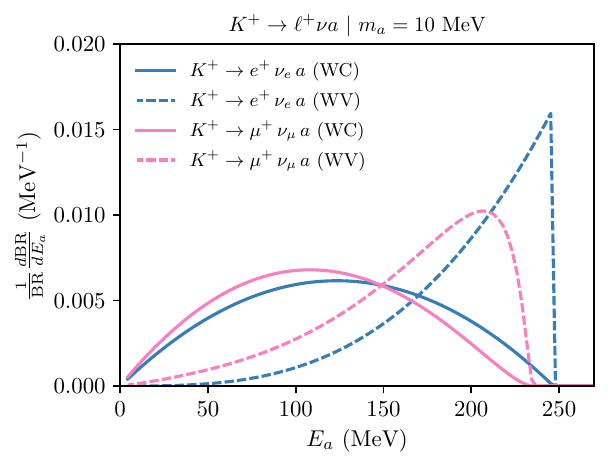}
    \caption{The normalized differential branching ratios for a $10$~MeV ALP in LFC models of the weak-conserving (left) and weak-violating (right) type.
    \label{fig:branchings_lalp_normalized}
    }
\end{figure*}

\begin{figure*}[t]
    \centering
    \includegraphics[width=\columnwidth]{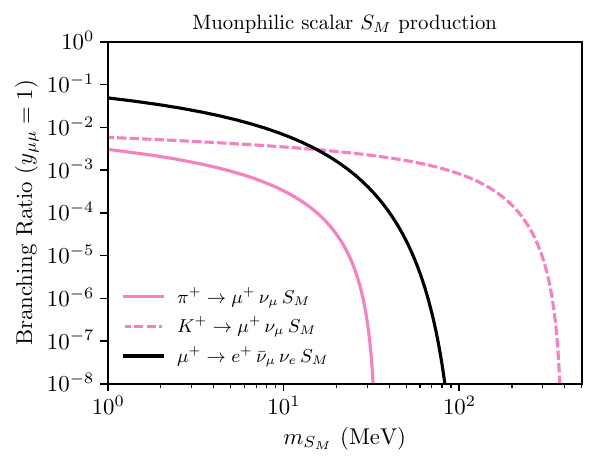}
    \includegraphics[width=\columnwidth]{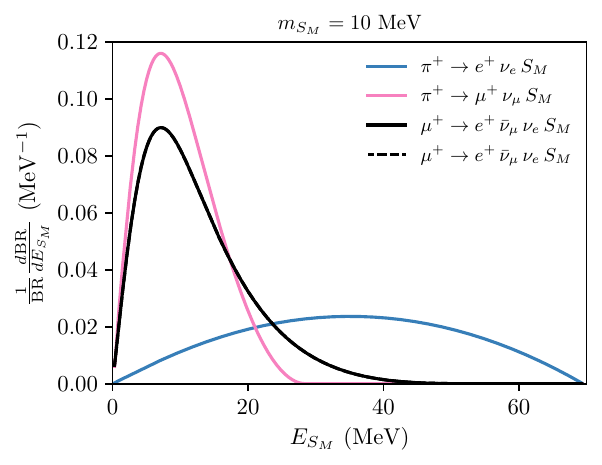}
    \caption{Left) The pion, kaon, and muon branching ratios into the muonphilic scalar $S_M$ as a function of the scalar mass.
    Right) the normalized differential branching ratios for a $10$~MeV scalar boson.
    \label{fig:branchings_Sm}
    }
\end{figure*}

In this appendix, we collect all three-body and four-body decay branching ratios for ALP and scalar boson production in pion, kaon, and muon decays.

\subsection{Three-body semi-leptonic meson decays}

Firstly, the three-body semi-leptonic decays of pseudo-scalar mesons $M=\pi,K$, namely
\begin{equation}
    M^+ \to \ell^+ \nu_\ell X.
\end{equation}
We present the double-differential branching ratio in the energy fractions of the new boson and the charge lepton, namely $x_a= 2 E_a/m_P$ and $x_\ell = 2 E_\ell/m_P$, and the mass ratios $r_i = m_i^2/m_P^2$.
Neglecting LFV couplings and structure-dependent contributions, we have
\begin{widetext}
\begin{align}
&\frac{\dd \mathcal{B}(P^+ \rightarrow \ell_i^+ \nu_i X)}{\dd x_\ell \dd x_a} \simeq \frac{\mathcal{B}(P^+ \rightarrow \ell_i^+ \nu_i)}{ (1 - r_\ell)^2} \frac{m_P^2}{m_{\ell_i}^2 16 \pi^2 f_a^2} \left[\frac{(c^L_{ii} + c^\nu_{ii})^2}{4} \left((x_a- 2 x_\ell x_a) - (1 - x_\ell)^2 - r_a + r_\ell\right)\right.
\\ \nonumber
&\left. + (c^L_{ii}-c^R_{ii})^2 r_a 
\frac{2 - 3 (x_a + x_\ell) + (x_a + x_\ell)^2 }{(1 - x_a - x_\ell + r_\ell)^2}
+
(c^R_{ii}-c^L_{ii}) (c^R_{ii}-c^L_{ii}-c^\nu_{ii})\frac{1 - r_a + r_\ell - (x_\ell + 2 x_a) + x_\ell x_a}{1 - x_a - x_\ell + r_\ell} \right].
\end{align}
\end{widetext}
The integration ranges are defined as $2\sqrt {r_a} < x_a  < 1 + r_a - r_\ell$ and $x_\ell^-< x_\ell < x_\ell^+$ where $x_\ell^\pm = {(2 - x_a)(1 + r^*) \pm \sqrt{x_a^2 - 4 r_a}(1-r^*)}$ and $r^*= m_\ell^2/(m_P^2 + m_a^2 - 2m_PE_a)$. We show the total branching ratios for the ALP LFC models in \cref{fig:branchings_lalp}, and the corresponding normalized differential branching ratios in \cref{fig:branchings_lalp_normalized}.

%%%%%%%%%%%%%%%%%%%%%%%%%%%%%%%%%%%%%%%%%%
\subsection{Four-body leptonic decays}\label{app:four-body-leptonic}

Now we tackle the four-body decay
\begin{equation}
    \mu^+ \to e^+ \nu_e \bar\nu_\mu X.
\end{equation}
We follow and extend the calculation of \cite{Ema:2025bww} to the general ALP couplings of \cref{eq:alp_leptonic_current} as well as to the MPS in \cref{eq:muonphilic}.
With these models in mind, the general form of the decay rate can be expressed as
\begin{align}
&\frac{\dd \Gamma(\ell_i^+ \rightarrow \bar\nu_i\nu_j \ell_j^+ X)}{\dd x} = \frac{G_F^2 m_{\ell_i}^5}{192 \pi^3} \frac{\hat I(\bar x)}{16\pi^2} \frac{\bar x F_{X}(\bar x)}{(1-\bar x)^2} \lambda^{1/2}(1, \bar x, r_X)
\end{align}
where $r_X = m_X^2/m_{\ell_i}^2$, $\bar x = (P_i-p_X)^2/m_{\ell_i}^2 = 1 + r_X - x$ is the dimensionless invariant mass of the virtual lepton $\ell_i$ and $x = 2 E_X/m_{\ell_i}$ is the energy fraction of the daughter boson.
The remaining factors include a universal function resulting from the integral over the daughter leptons,
\begin{widetext}
\begin{align}
    \hat I(\bar x) &= \int_{0}^{(\sqrt{\bar x} - \sqrt{r_\ell})^2} \left(1 + \frac{x_{\nu\nu} - 2 r_\ell}{\bar x} + \frac{r_\ell^2 + r_\ell x_{\nu\nu} - 2 x_{\nu\nu}^2}{\bar x^2}\right)\lambda^{1/2}\left(1, \frac{x_{\nu\nu}}{\bar x}, \frac{r_\ell}{\bar x}\right)\dd x_{\nu\nu}
    \\\nonumber
    &=\frac{1}{2}\left(
        \bar x
        - 8 r_\ell
        + 12 \frac{r_\ell^2}{\bar x} \log\left(\frac{\bar x}{r_\ell}\right)
        + 8 \frac{r_\ell^3}{\bar x^2}
        - \frac{r_\ell^4}{\bar x^3}
    \right),
\end{align}
where $x_{\nu\nu} = (p_{\nu_i}+p_{\nu_j})^2/m_{\ell_i}^2$ and $\lambda(x,y,z) = (x-y-z)^2 - 4 y z$.

The function $F_X(\bar x)$ carries the model dependence of the amplitude.
For the ALP models, it is given by
\begin{align}
    F_{a_\ell} (\bar x) = \frac{m_{\ell_i}^2}{4 f_a^2} \Bigg[&
    \left(\bar x (c^L_{ii})^2 + \, (c^R_{ii})^2\right)\left((1-\bar x)^2- r_a (1+\bar x)\right) + 4 c^L_{ii} c^R_{ii}r_a \bar x
    \\ \nonumber
    &+ (c^\nu_{ii})^2(1-\bar x)^2(1 + \bar x - r_a) 
    - 2 c^L_{ii}c^\nu_{ii}\bar x (1 - \bar x) (1 - \bar x + r_a)
    - 2 c^R_{ii}c^\nu_{ii}\bar x (1 - \bar x) (1 - \bar x - r_a)
\Bigg].
\end{align}
\end{widetext}
In the WC regime ($c^\nu_{ii} = c^L_{ii}$, $c^L_{ii}=-c^R_{ii}=-1$),
\begin{align}
    F_{a_\ell}^{\rm WC} (\bar x) &= \frac{m_{\ell_i}^2}{f_a^2} \left[(1-\bar x)^2  - r_a (1+\bar x)\right],
\end{align}
while in the WV scenario ($c^\nu_{ii} = 0$, $c^L_{ii}=-c^R_{ii}=-1$), we have
\begin{align}
    F_{a_\ell}^{\rm WV} (\bar x) &= \frac{m_{\ell_i}^2}{4f_a^2} \left[(1-\bar x)^2 (1+\bar x) -r_a (\bar x (\bar x + 6 ) + 1)\right],
\end{align}
Contrary to the case of meson decays, both amplitudes are proportional to the lepton mass.
The difference between WV and WC cases in fully leptonic decays will be most pronounced in the energy spectrum of $a_\ell$, rather than in the total rates.

For the muonphilic scalar of \cref{eq:muonphilic}, we find
\begin{align}
    F_{S_M} (\bar x) = \left[ (c_L^2 + \bar x c_R^2) (1 + \bar x - r_a)  + 4 \bar x c_L c_R\right],
\end{align}
with $c_{L,R} = (c_S\pm c_P)/2$.

In \cref{fig:branchings_lalp,fig:branchings_lalp_normalized,fig:branchings_Sm} we show the total and differential branching ratio of muons into the three different types of bosons considered here.

%%%%%%%%%%%%%%%%%%%%%%%%%%%%%%%%%%%%%%%%%%%%%%%%%%
\bibliography{refs}
\bibliographystyle{apsrev4-1}

%%%%%%%%%%%%%%%%%%%%%%%%%%%%%%%%%%%%%%%%%%%%%%%%%%

\end{document}